\documentclass[fleqn,usenatbib]{mnras}
\usepackage{newtxtext,newtxmath}
\usepackage[T1]{fontenc}

\DeclareRobustCommand{\VAN}[3]{#2}
\let\VANthebibliography\thebibliography
\def\thebibliography{\DeclareRobustCommand{\VAN}[3]{##3}\VANthebibliography}

\usepackage{graphicx}

\usepackage{amsmath}
\usepackage{amssymb}

\newcommand{\vpeak}{V_{\rm peak}}
\newcommand{\vmax}{V_{\rm max}}

\newcommand{\sigmaLogM}{\sigma_{\rm logM}} 
\newcommand{\Mone}{M_{\rm 1}}
\newcommand{\Mmin}{M_{\rm min}} 
\newcommand{\Mcut}{M_{\rm cut}} 

\newcommand{\sigL}{\sigma_{\rm lum}}
\newcommand{\tmerger}{t_{\rm merger}}

\newcommand{\FkP}{f_{\rm k,cen+sat}}
\newcommand{\FkM}{f_{\rm k,cen-sat}}
\newcommand{\betaL}{\beta_{\rm lum}}

\newcommand{\hMsun}{ h^{-1}{\rm M_{ \odot}}}
\newcommand{\hMpc}{ h^{-1}{\rm Mpc}}

\newcommand{\ihMpcC}{ h^{3}{\rm Mpc}^{-3}}

\newcommand{\sig}{\sigma_{8}}
\newcommand{\OmM}{\Omega_\mathrm{M}}
\newcommand{\Omb}{\Omega_{\rm b}}
\newcommand{\h}{h}
\newcommand{\ns}{{n_{\rm s}}}
\newcommand{\Mnu}{M_{\rm \nu}}
\newcommand{\wa}{w_{\rm a}}
\newcommand{\wz}{w_{0}}

\newcommand{\proj}{${w_{\rm p}}$}
\newcommand{\mono}{$\xi_{\ell=0}$}
\newcommand{\quadr}{$\xi_{\ell=2}$}

\newcommand{\lensing}{$\Delta \Sigma$}

\newcommand{\shame}{SHAMe~}

\newcommand{\Planck}{Planck}
\newcommand{\LowS}{Low S8}

\title[Reproducing the GC and GGL of BOSS]{Consistent clustering and lensing of SDSS-III BOSS galaxies with an extended abundance matching formalism
}

\author[S. Contreras et al.]{
Sergio Contreras$^{1}$\thanks{E-mail: \href{mailto:sergio.contreras@dipc.org}{sergio.contreras@dipc.org}},
Jon\'{a}s Chaves-Montero$^{2}$\thanks{E-mail: \href{mailto:jchaves@ifae.es}{jchaves@ifae.es}} and
Raul E. Angulo$^{1,3}$\thanks{E-mail: \href{mailto:reangulo@dipc.org}{reangulo@dipc.org}}
\\
$^{1}$ Donostia International Physics Center, Manuel Lardizabal Ibilbidea, 4, 20018 Donostia, Gipuzkoa, Spain\\
$^{2}$ Institut de F\'isica d'Altes Energies, The Barcelona Institute of Science and Technology, Campus UAB, E-08193 Bellaterra (Barcelona), Spain.\\
$^{3}$ IKERBASQUE, Basque Foundation for Science, 48013, Bilbao, Spain.\\
}

\date{Accepted XXX. Received YYY; in original form ZZZ}

\pubyear{2023}

\begin{document}
\label{firstpage}
\pagerange{\pageref{firstpage}--\pageref{lastpage}}
\maketitle

\begin{abstract}
Several analyses have shown that $\Lambda$CDM-based models cannot jointly describe the clustering (GC) and galaxy-galaxy lensing (GGL) of galaxies in the SDSS-III BOSS survey, which is commonly known as the ``lensing-is-low problem''. In this work, we show that an extension of Subhalo Abundance Matching, dubbed SHAMe, successfully solves this problem. First, we show that this model accurately reproduces the GC and GGL of a mock galaxy sample in the TNG300 hydrodynamic simulation with properties analogous to those of BOSS galaxies. Then, we switch our attention to observed BOSS galaxies at $z=0.31-0.43$, and we attempt to reproduce their GC and GGL by evaluating SHAMe on two different simulations: one adopting best-fitting cosmological parameters from \Planck~and the other from weak gravitational lensing surveys (\LowS), where the amplitude of matter fluctuations is lower for the latter. We find excellent agreement between SHAMe predictions and observations for both cosmologies, indicating that the lensing-is-low problem originates from approximations in previous theoretical descriptions of the data. The main difference between SHAMe results in these cosmologies is the level of galaxy assembly bias, which is approximately 20 and 10\% for \Planck~and \LowS, respectively. These results highlight the dangers of employing oversimplified models to analyse current large-scale structure datasets, and the need for realistic yet flexible descriptions of the galaxy-halo connection.
%We check that our findings also hold for other higher and lower redshift samples of the BOSS survey.
\end{abstract}

\begin{keywords}
large-scale structure of Universe --- gravitational lensing: weak --- galaxies: statistics --- galaxies: haloes --- cosmology: observations --- cosmology: theory
\end{keywords}

\section{Introduction}

In the modern galaxy formation formalism, galaxies form and evolve within dark matter haloes. As haloes evolve hierarchically, merging together to form the largest structures in the universe, galaxies within the haloes begin to cluster and ultimately merge. While the rate at which haloes merge is only governed by the cosmological model, the merging of galaxies is also influenced by baryonic processes. Therefore, the large-scale distribution of galaxies encodes precise information about cosmology and galaxy formation physics, but disentangling this information is challenging.

Two of the observables most widely used to extract this information are the projected correlation function (\proj) and galaxy-galaxy lensing (GGL). The first captures the excess probability of finding a pair of galaxies as a function of projected distance, while the second measures the amount by which foreground galaxies distort the light of background galaxies surrounding these \citep[e.g.,][]{Tyson:1984, Miralda:1991, Brainerd:1996, Hudson:1998}. Due to integrating line-of-sight information, these are not significantly affected by either redshift space distortions or uncertainties affecting galaxy redshifts, making them ideal for photometric redshift surveys.

Traditionally, the projected correlation function has been used on its own to set constraints on the average occupation of galaxies as a function of halo mass (e.g., \citealt{Zehavi:2005, Zheng:2007}), galaxy assembly bias (e.g., \citealt{Salcedo:2022}), and the value of cosmological parameters (e.g., \citealt{AEMULUS3, AEMULUS5}) using empirical models such as the halo occupation distribution (HOD, \citealt{Jing:1998a, Benson:2000, Peacock:2000, Berlind:2003, Zheng:2005, Zheng:2007, C13, Guo:2015a, C17}) or the subhalo abundance matching technique (SHAM, \citealt{Vale:2006, Conroy:2006}). More recently, when trying to carry out joint analyses of galaxy clustering (GC) and GGL, multiple studies have found a systematic difference between GGL observations and predictions from empirical models. This effect, now referred to as the lensing-is-low problem, was first detected by \cite{Leauthaud:2017} when using HOD and SHAM models to analyse the CMASS galaxy sample of the Baryon Oscillation Spectroscopic Survey \citep[BOSS,][]{Eisenstein:2011, Dawson:2013} and lensing measurements around these galaxies from the Canada France Hawaii Telescope (CFHT) Lensing Survey \citep[CFHTLenS,][]{Heymans:2012, Miller:2013} and CFHT Stripe 82 survey \citep{Erben:2013}. This study found that the observed GGL signal is from 20 to 40\% weaker than predictions from analytical models. Later, \cite{Lange:2019b} confirmed these results by examining additional BOSS galaxy samples, this time only using HODs to model galaxy samples.

Currently, there is no complete consensus on the level of the lensing-is-low effect, with large discrepancies between various studies \citep{Wibking:2020, Yuan:2020, Yuan:2021, Lange:2021b, Yuan:2022b, Yuan:2022, Lange:2023} despite using similar (if not identical) galaxy samples and implementations of HOD models. \cite{Amon:2023} carried out the most statistically significant analysis so far, finding 20-30\% small-scale discrepancy that decreases with increasing scale when using a HOD model to reproduce the projected correlation function of BOSS galaxies and GGL measurements for these galaxies from the Dark Energy Survey year 3 data release \citep{Amon:2022}, the fourth Kilo-Degree Survey data release  \citep{Asgari:2021}, and the Subaru Hyper Suprime-Cam survey year 1 data release \citep{Hikage:2019}. 

In \cite{Chaves:2023}, we explored the origin of the lensing-is-low problem using a sample of galaxies from the TNG300 hydrodynamic simulation (e.g., \citealt{TNGa, TNGb}) with analogous properties as the low-redshift sample of the BOSS survey \citep[LOWZ;][]{Eisenstein:2011}. Traditionally, the lensing-is-low problem refers to the fact that empirical models optimised to reproduce the small-scale clustering of BOSS galaxies overpredict their galaxy-galaxy lensing by about 30\%. In this work, we broad the scope of this definition to a similar problem for any type of galaxy sample from observations or simulations. Using a standard HOD model, we found a lensing-is-low problem for this sample with the same amplitude and scale dependence as the one found in observations (e.g., \citealt{Lange:2019b, Amon:2023}). We then investigated the origin of this problem, finding that resulted from ignoring a variety of galaxy formation effects in standard HOD models, including assembly bias, segregation of satellite galaxies relative to dark matter, and baryonic effects on the matter distribution. This result suggests that one of the most popular solutions to the lensing-is-low problem, which states that it is another face of the tension between the growth of structure measurements from the early and late Universe \citep{Leauthaud:2017, Lange:2019, Wibking:2020,  Yuan:2022, Amon:2023}, is likely incorrect or at least incomplete.

Motivated by these findings, in \cite{C23a} we examined the ability of the SubHalo Abundance Matching extended model (SHAMe, \citealt{C21c}), an empirical model with a richer implementation of galaxy formation effects than most HOD models, to reproduce GC and GGL jointly. We showed that this model successfully captures the projected correlation function, monopole and quadrupole of the correlation function, and galaxy-galaxy lensing signal of $M_r$-selected galaxies from the TNG300 simulation. In this work, we first check whether SHAMe can also jointly reproduce the GC and GGL of the colour-selected sample of BOSS-like galaxies from the TNG300 simulation used in \citet{Chaves:2023}. Then, we focus our attention on galaxies from the LOWZ sample of the BOSS survey, with the aim of reproducing their GC and GGL as well as confirming whether the origin of the lensing-is-low problem is astrophysical. To do so, we evaluate SHAMe on two different cosmological gravity-only simulations: the first presents best-fitting cosmological parameters from \textit{Planck} (``\Planck'' cosmology) and the second from weak lensing surveys (``\LowS'' cosmology), which has been claimed to ease the lensing-is-low problem (e.g., \citealt{Amon:2023}). We successfully recover the GC and GGL of observed galaxies for both cosmologies, thereby confirming that the origin of the lensing-is-low problem is not necessarily cosmological. Then, we show that the main difference between SHAMe predictions for both cosmologies is the level of galaxy assembly bias, thereby confirming the solution proposed by \cite{Chaves:2023}. In addition to resolving the lensing-is-low problem, this work shows that the SHAMe model can be used to create mocks that consistently reproduce both galaxy clustering (in real and redshift space) and galaxy-galaxy lensing.

The outline of this paper is as follows: Section~\ref{sec:simulations} presents the dark matter simulations and galaxy population models used in this work. The computation of the galaxy clustering and the quantification of errors in our model are presented in Section~\ref{sec:methodology}. We test the performance of our methodology on a hydrodynamic simulation in Section~\ref{sec:cluster_TNG}. The main results of this paper, the measurement of the galaxy clustering and galaxy-galaxy lensing are presented in Section~\ref{sec:cluster_BOSS}. Additional statistics inferred from the clustering, such as the galaxy assembly bias and the HOD are shown in Section~\ref{sec:HOD_GAB}. We finalize by presenting our conclusions in Section~\ref{sec:fin}.

Unless otherwise stated, the standard units in this paper are $\hMsun$ for masses and $\hMpc$ for distances. All logarithm values are in base 10.

\section{Numerical simulations \& galaxy samples}
\label{sec:simulations}

In this section, we first describe the empirical model SHAMe (\S~\ref{sec:shame}) and the simulations we populate using it (\S~\ref{sec:DMonly}). Then, we introduce the samples that we analyse using SHAMe: observed galaxies from the BOSS survey (\S~\ref{sec:obs}) and simulated galaxies from the TNG300 simulation (\S~\ref{sec:tng}).

\subsection{Sub-halo abundance matching extended model}
\label{sec:shame}

To reproduce the GC and GGL of a galaxy sample, we need a model capable of populating dark matter simulations realistically and efficiently. To accomplish this, we employ the {\bf S}ub-{\bf H}alo {\bf A}bundance {\bf M}atching {\bf e}xtended model (SHAMe) developed by \cite{C21a, C21c}. The two primary advantages of this model are (a) the small number of free parameters compared to other models such as HODs and (b) its ability to reproduce the GC and GGL of a galaxy sample simultaneously. We showed this last feature in \cite{C23a}, where we fit the projected correlation function, the monopole and quadrupole of the correlation function, and the galaxy-galaxy lensing signal of $\rm M_r-$selected galaxies. Additionally, the model is also capable of reproducing galaxy clustering with redshift space distortion at large and small scales.

Just as in the standard SHAM approach \citep{Vale:2006, Conroy:2006, Reddick:2013, C15, ChavesMontero:2016, Lehmann:2017, Dragomir:2018,  Hadzhiyska:2021b, Favole:2022}, our model begins by matching a subhalo property (in this case, $\vpeak$) to the expected luminosity function. For a colour-selected galaxy sample like this one, this only fulfils the purpose to assign a physical unit to the value of the scatter of the SHAMe model, since as is mentioned in \cite{C21c}, when galaxies are selected using number density cuts, the choice of luminosity function has no significant effect on galaxy clustering statistics. This test was repeated for the GGL signal reaching the same conclusion. In this paper, we used the luminosity function in $M_r$ from \citet{Blanton:2001} for SDSS at z=0.1.

After creating this mapping, the model introduces orphan galaxies by assuming that these merge with its central structure when the time since accretion exceeds a dynamical friction timescale, $t_{\rm infall} > t_{\rm d.f.}$, where $t_{\rm d.f.}$ is the dynamical friction time computed at the time the satellite subhalo became an orphan using a modified version of Eq. 7.26.\cite{BT:1987},
\begin{equation}
t_{\rm d.f.} = \dfrac{1.17\ \tmerger\ d_{\rm host}^2\ V_{\rm host} (M_{\rm host}/10^{13}\ h^{-1}\mathrm{M}_{\odot})^{1/2}}{G \ln(M_{\rm host}/M_{\rm sub}+1)\ M_{\rm sub}},
\end{equation}
\noindent where $\tmerger$ is a dimensionless free parameter that effectively regulates the number of orphan galaxies; $d_{\rm host}$ is the distance of the subhalo to the centre of its host halo; $V_{\rm host}$ is the virial velocity of the host halo; $M_{\rm host}$ is the virial mass of the host halo, and $M_{\rm sub}$ is the subhalo mass. 

SHAMe also presents an option for removing galaxies that became satellites a long time ago, which is justified because satellite galaxies progressively lose cold gas, stellar mass, and luminosity over time. We find we can improve galaxy clustering predictions by excluding galaxies with $t_{\rm infall} > \betaL\, t_{\rm dyn}$, with $t_{\rm dyn}$ the halo's dynamical time, defined as $0.1/H(z)$ and $\betaL$ being a free parameter. We also tested alternative methods, such as removing substructures based on their lost subhalo mass (as in \citealt{Moster:2018} and \citealt{C21a}) and other more complex methods, and found that their performance was equivalent or inferior.

As the final step in the SHAMe implementation, additional galaxy assembly bias is incorporated. Galaxy assembly bias \citep{Croton:2007} is the change in galaxy clustering caused by the propagation of halo assembly bias into the galaxies. This propagation occurs because the occupation of galaxies depends on halo properties that cause halo assembly bias (i.e.,~occupancy variations, \citealt{Zehavi:2018, Artale:2018}). As far as we are aware, there has been no definitive confirmation of the (non)existence of this type of assembly bias for observed galaxies, and the level of assembly bias is not the same for SHAM and hydrodynamical simulations \citep{ChavesMontero:2016, Hadzhiyska:2022a, Hadzhiyska:2022b} or semi-analytic models \citep{C21c, Hadzhiyska:2021}. In addition, none of these necessarily corresponds to the level of assembly bias in the actual universe. To account for the uncertainty surrounding the assembly bias of the target galaxy sample, we introduce a tunable bias level into our galaxy model.

We introduce variable assembly bias in our model using the method developed by \cite{C21a}, which employs the individual bias-per-object of the galaxies \citep{Paranjape:2018} to select preferentially more/less biased objects. This individual bias-per-object estimate corresponds to the cross-correlation between a given point in space and the dark matter density field and is measured by computing the power spectrum around each object for scales between $0.08 < {\rm k}\ h\ {\rm Mpc^{-1}} < 0.316$. In essence, the model swaps the luminosities of galaxies with similar $\vpeak$ values so that their luminosities correlate/anticorrelate with large-scale environment density (see also \citealt{Hadzhiyska:2020, Xu:2021b, Xu:2021a} for other studies that look at the impact of environment on other galaxy population models). By performing this step independently for central and satellite galaxies, we preserve the satellite fraction of the original galaxy sample. To control the level of galaxy assembly bias, SHAMe uses $f_{\rm k,cen}$ and $f_{\rm k, sat}$ for central and satellite galaxies, respectively. A value of $f_{\rm k}=1$ (-1) results in the maximum positive (negative) galaxy assembly bias signal, while a value of 0 leads to the same level of assembly bias as a standard SHAM. For simplicity, we express the assembly bias parameters as $f_{\rm k,cen+sat} = f_{\rm k,cen}+f_{\rm k, sat}$ and $f_{\rm k,cen-sat} = f_{\rm k,cen}-f_{\rm k, sat}$, where a value of $f_{\rm k,cen-sat} = 0$ represent the same level of assembly bias for centrals and satellites. It is important to highlight that, since the galaxy assembly bias is added by exchanging the luminosity of the galaxies in bins of $\vpeak$, a low or null value of the parameter controlling the scatter between these properties results in a negligible amount of assembly bias even for $f_{\rm k}=1$ (-1).

\subsection{The gravity-only simulations}
\label{sec:DMonly}

\begin{table}
    \centering
        \caption{The cosmological parameters of the two main pairs of simulations used in this work. Both simulations have values of $\Mnu=0$, $\wz=-1$, and $\wa=0$.}
    \begin{tabular}{ccccccccccc}
        \hline
        Cosmology &  $\sig$ &  $\OmM$ & $\Omb$ & $\h$ & $\ns$ & $\rm S_8$\\
        \hline
        \Planck  & 0.810 & 0.310 & 0.049 & 0.677 & 0.967 & 0.823\\
        \LowS  & 0.754 & 0.305 & 0.047 & 0.682 & 0.965 & 0.760\\
        \hline
    \end{tabular}
    \label{table:sims}
\end{table}

To build our mocks, we run two pairs of gravity-only simulations with different cosmologies, one with a ``\Planck~cosmology'', and one with a Planck cosmology with a lower value of $S_8$, similar to the ``Lensing Cosmology'' from \cite{Amon:2023}. We gather the cosmological parameters adopted by these simulations in Table~\ref{table:sims}. Each pair was run with opposite initial Fourier phases, using the procedure of \cite{Angulo:2016} that suppresses cosmic variance by up to 50 times compared to a simulation with random phases for the same volume. Each simulation has a volume of $(512 \hMpc)^3$, a resolution of $1536^3$ particles, and were carried out using an updated version of {\tt L-Gadget3} \citep{Angulo:2012}, a lean version of {\tt GADGET} \citep{Springel:2005} used to run the Millennium XXL simulation \citep{Angulo:2012} and the Bacco Simulations \citep{Angulo:2021}. This implementation enables on-the-fly identification of haloes and subhaloes using a Friend-of-Friend algorithm \citep[{\tt FOF}][]{Davis:1985} and uses an extended version of {\tt SUBFIND} \citep{Springel:2001} that better identifies substructures by considering their past history. In addition, this version of {\tt SUBFIND} provides properties that are non-local in time, such as the maximum circular velocity ($\vmax \equiv {\rm max}\sqrt{GM(<r)/r}$) attained by a halo/subhalo during its evolution the maximum circular velocity, $\vpeak$, and is capable to identify orphan substructures, i.e.~satellite structures with known progenitors that the simulation cannot resolve but that are expected to exist in the halo, by tracing the most bound particle of all subhalos after these cannot be identified.

To facilitate the comparison with the TNG300, we carried out a gravity-only simulation with the same cosmology, volume, and initial conditions as its hydrodynamic counterpart, but with a reduced resolution ($625^3$ particles). We refer to this simulation as ``TNG300-mimic'' (see \citealt{C23a} for a detailed description of this simulation). 

\subsection{Observational data}
\label{sec:obs} 

We use galaxy clustering measurements from the BOSS survey and the GGL measurements around BOSS galaxies from the Dark Energy Survey year 3 data release \citep{Amon:2022, Secco:2022}, the fourth Kilo-Degree Survey data release \citep{Asgari:2021}, and the Subaru Hyper Suprime-Cam survey year 1 data release \citep{Hikage:2019}, all these presented and provided by \cite{Amon:2023}. We concentrate specifically on LOWZ galaxies at $z=0.31-0.43$ because GGL measurements from different surveys show great consistency for this sample \citep{Leauthaud:2022, Lange:2023}. Nonetheless, we also examine the remaining samples analysed by Amon et al. (i.e., LOWZ $z = 0.15 - 0.31$, CMASS $z=0.43-0.54$ \& CMASS $z=0.54-0.70$) and find results comparable to those presented in this paper (see Appendix~\ref{sec:Ap1} for a summary of the performance of the SHAMe model on these samples). Following \cite{Amon:2023}, we build our mocks at the median redshift of each of these samples to compare with the observational data (z = 0.24 \& 0.364 for the LOWZ samples; z=0.496 \& 0.592 for the CMASS samples) and assuming a number density of galaxies of $n=3.26$ and $3.01\times 10^{-4}\ihMpcC$ for LOWZ and CMASS, respectively.

\subsection{TNG300 simulation}
\label{sec:tng}

We study the performance of \shame jointly reproducing GC and GGL using galaxies from the Illustris-TNG300 simulation\footnote{\url{https://www.tng-project.org/}} (thereafter TNG300, \citealt{TNGa, TNGb, TNGc, TNGd, TNGe}), a successor of the Illustris simulation \citep{Illustrisa,Illustrisb,Illustrisc,Illustrisd}. TNG300 solves for the joint evolution of dark matter and baryons in a periodic box of 205 $\hMpc$ ($\sim300$ Mpc) on a side using $2500^3$ dark matter particles and gas cells, implying a baryonic mass resolution of $7.44\times10^6\,\hMsun$ and a dark matter particle mass of $3.98\times 10^7\,\hMsun$. As a result, TNG300 is one of the largest publicly available, high-resolution hydrodynamic simulations. This simulation adopted cosmological parameters consistent with recent analyses \citep{Planck2015}.

As shown in \cite{Springel:2018}, the clustering of TNG galaxies agrees well with observational estimates across a wide range of stellar masses. In addition, as demonstrated by \cite{Renneby:2020}, the GGL of TNG galaxies agrees well with KIDS+GAMA and SDSS observations, including for galaxies selected according to stellar mass, colour, group membership, and isolation. 

Specifically, we use the BOSS-TNG galaxy sample of the TNG300, a selection of galaxies made by \cite{Chaves:2023} which reproduces the number density and colour selection of the LOWZ galaxies. When fitting the clustering of this sample using a standard HOD model, \cite{Chaves:2023} found a lensing-is-low problem with similar amplitude and scale dependence as the one found in observational studies (e.g., \citealt{Lange:2019b, Amon:2023}). Therefore, this sample is an ideal testbed for SHAMe.

\section{Methodology}
\label{sec:methodology}

In this section, we describe the computation of GC and GGL (\S\ref{sec:lensing}), the construction of an emulator to speed up predictions for these observables (\S\ref{sec:emu}), and the algorithms used to find the best-fitting solution of SHAMe to GC and GGL measurements (\S~\ref{sec:PSO_MCMC}).

\subsection{Clustering \& lensing}
\label{sec:lensing}

\subsubsection*{Galaxy clustering}

The GC statistics we use in this work are the projected correlation function (\proj) and the monopole (\mono) and the quadrupole (\quadr) of the correlation function (the latter two just for TNG300 data). We utilised \textsc{corrfunc} \citep{Corrfunc1, Corrfunc2} to compute the projected correlation function, which is obtained by integrating the 2-point correlation function, $ \xi(r_{\uppi}, r_\mathrm{p})$, over the line-of-sight, 
\begin{equation}
    w_{\rm p} = 2 \int_{0}^{\pi_{\rm max}} \xi( r_{\uppi}, r_\mathrm{p}) \mathrm{d} r_{\uppi},
\end{equation}
\noindent with $\pi_{\rm max}=30$ and $100\ \hMpc$ the maximum depth used for the TNG300 and BOSS, respectively. We do not consider larger scales on the TNG300 because the results become too noisy due to the simulation's small volume.

To measure the multipoles of the correlation function, we first measure the 2-point correlation function in bins of $s$ and $\mu$, where $s^2 = r^2_{\rm p} + r^2_{\uppi}$ and $\mu$ is the cosine of the angle between $s$ and the line-of-sight. We compute this statistic once again with \textsc{corrfunc}. The correlation functions for multipoles are then defined as follows: 
\begin{equation}
    \xi_{\ell} = \frac{2 \ell+1}{2} \int^{1}_{-1} \xi(s,\mu)P_\ell(\mu)d\mu
\end{equation}
\noindent with $P_{\ell}$ is the $\ell$-th order Legendre polynomial.

\subsubsection*{Galaxy-galaxy lensing}

A foreground mass distribution induces a shear signal on background sources that varies with the lens-source pair's separation. The signal stacked from a sufficient number of lenses is proportional to the excess surface density
\begin{equation}
    \Delta\Sigma(r_\perp) = \overline{\Sigma}(\leq r_\perp) - \Sigma(r_\perp),
\end{equation}
where $\Sigma$ is the azimuthally-averaged surface mass density and $\overline{\Sigma}(\leq r_\perp)$ is the mean surface density within projected radius $r_\perp$. We estimate the azimuthally-averaged surface mass density using
\begin{equation}
    \Sigma(r_\perp) = \Omega_\mathrm{m} \rho_\mathrm{crit} 
    \int_{-r_\parallel^\mathrm{max}}^{r_\parallel^\mathrm{max}} \xi_\mathrm{gm}(r_\perp, r_\parallel)\,\mathrm{d}r_\parallel,
\end{equation}
where $r_\parallel$ and $r_\perp$ refer to the projected distance along and perpendicular to the line-of-sight from the lens galaxy, $r_\parallel^\mathrm{max}$ is the integration boundary, $\xi_\mathrm{gm}$ is the galaxy-matter three-dimensional cross-correlation, $\Omega_\mathrm{m}$ and $\rho_\mathrm{crit}$ are the matter and critical density of the Universe, respectively, and the mean surface mass density within a radius $r_\perp$ is
\begin{equation}
    \overline{\Sigma}(\leq r_\perp) = \frac{2}{r_\perp^2}\int_0^{r_\perp} \Sigma(\tilde{r})\,\tilde{r} \,\mathrm{d}\tilde{r}.
\end{equation}

We measure $\xi_{\rm gm}$ using again \textsc{corrfunc}, with $r_\parallel^\mathrm{max}=30$ and $100\ \hMpc$ for TNG300 and BOSS, respectively, and a subsampled version of matter density field diluted by a factor of $\sim 3000$. In \cite{C23a}, we verified that this dilution factor ensures sub-percent precision for $\Delta\Sigma$ measurements across the entire range of scales considered from our simulations. In order to reduce uncertainty in our calculations, we compute the mean of each observable after considering the line-of-sight to be each of the three simulation coordinate axes.

\subsection{Emulator}
\label{sec:emu}

One of the main advantages of the SHAMe model is its performance. Using a single CPU, it takes less than 1.5 minutes to populate a $(512 \hMpc)^3$ dark matter simulation. Due to the efficiency of \textsc{corrfunc}, the clustering measurements: 3 projected correlation functions and 3 galaxy-galaxy lensing measurements (one per line-of-sight) that integrate to up to $100 \hMpc$ only take 2.7 minutes for such simulation using a single CPU.

While reasonably fast, running an MCMC or another similar technique, such as a Bayesian analysis or an iterative emulator (e.g., \citealt{Pellejero-Ibanez:2020}), would still require a significant amount of time. To improve the computational efficiency, we construct an emulator that predicts GC and GGL in a fraction of a second. We construct a different emulator for the ``\Planck'' and ``\LowS'' cosmologies; to do so, we evaluate SHAMe on these simulations for $\simeq 10.000$ of its free parameters within the ranges
\begin{eqnarray}
\label{eq:par_range}
\sigL                 &\in& [0, 2],\\ 
\log(\tmerger)        &\in& [-1.5, 1.2],\\
\FkP                  &\in& [-0.5, 0.5],\\
\FkM                  &\in& [-0.5, 0.5],\\
\betaL t_{\rm dyn}    &\in& [0,9].
\end{eqnarray}
Then, we extract GC and GGL measurements for each BOSS sample from each simulation. To reduce discreteness errors on these observables, we evaluate SHAMe twice for each combination of parameters.

Similar to \cite{Angulo:2021, Arico:2021a, Arico:2021b, Pellejero:2022b, Zennaro:2021b, C23b}, we construct our emulators using a feed-forward Neural Network. The architecture consists of two fully-connected hidden layers with 200 neurons each and a Rectified Linear Unit activation function for the projected correlation function, as well as three fully connected and three layers for the GGL signal, with a separate network for each statistic. We explored alternative configurations obtaining comparable performances.

The neural networks were trained using the Keras interface of the Tensor-flow library \citep{tensorflow}. We implemented the Adam optimisation algorithm with a learning rate of 0.001 and a mean-squared error loss function. Our dataset is divided into distinct groups for training and validation purposes. A single Nvidia Quadro RTX 8000 GPU card required approximately 30 minutes per number density/statistic to process the training set, which contains 90\% of the data. On an 11-year-old Intel i7-3820, evaluating the two emulators takes $\sim 6$ milliseconds, while evaluating batches of 1,000,000 samples takes $\sim 22$ seconds (it is more efficient to evaluate the data in larger groups).

\subsection{MCMC \& PSO}
\label{sec:PSO_MCMC}

We use two different techniques to find the best-fitting parameters of SHAMe: a Markov chain Monte Carlo (MCMC) and a Particle Swarm Optimization (PSO).

We use the MCMC technique to compute the best-fitting solution of SHAMe to the GC and GGL of BOSS galaxies. Specifically, we employ the Affine Invariant Markov chain Monte Carlo Ensemble sampler {\tt emcee} \citep{emcee}\footnote{https://emcee.readthedocs.io/en/stable/}. For each step of the Markov chain, {\tt emcee} draws a value for each SHAMe parameter, and we estimate the probability of these by comparing BOSS measurements and SHAMe emulator predictions. We compute the likelihood function for the MCMC as $L = \chi^2/2$ with 
\begin{equation}
\label{eq:chi2}
\chi^2 = (V_{\rm target}-V_{\rm SHAMe})^T C_v^{-1} (V_{\rm target}-V_{\rm SHAMe}),
\end{equation}
with $V$ the GC or GC+GGL joining vector and $C_v$ the covariance matrix, which we assume diagonal and equivalent to the square of the observational uncertainties provided by \cite{Amon:2023}. We ran 1,000 chains of 20,000 steps each and a burn-in phase of 1,000 steps. This combination of chains and steps is ideal for an emulator-based MCMC, which is highly efficient when computing a large number of points concurrently. Additional MCMC configurations were evaluated, with nearly identical results. Each MCMC chain on a single CPU takes an average of one hour to compute.

We chose the model that better reproduces the target sample as the one with the highest (lowest) likelihood ($\chi^2$). We also take the 3.7\% of the elements of the MCMC chains (after the burn-in) with the highest likelihood as an estimator of the dispersion of the SHAMe mocks that have an overall good performance fitting the target GC and GGL. This selection is equivalent to the volume inside 1-$\sigma$ in a 5-dimension Gaussian, where this dimension is equal to the number of free parameters in SHAMe.

To find the best parameters that reproduce the GC and GGL of the TNG300 simulation we use the PSO algorithm PSOBACCO\footnote{\url{https://github.com/hantke/pso_bacco}} \citep[see][]{Arico:2021a}. This algorithm is more efficient to find the parameters that minimize the $\chi^2$ compared to an MCMC, which is the only thing we need when reproducing this simulation. Same as for the MCMC, we compute the $\chi^2$ using Eq.~\ref{eq:chi2}.

\section{The TNG300 test case}
\label{sec:cluster_TNG}

\begin{figure*}
\includegraphics[width=1.0\textwidth]{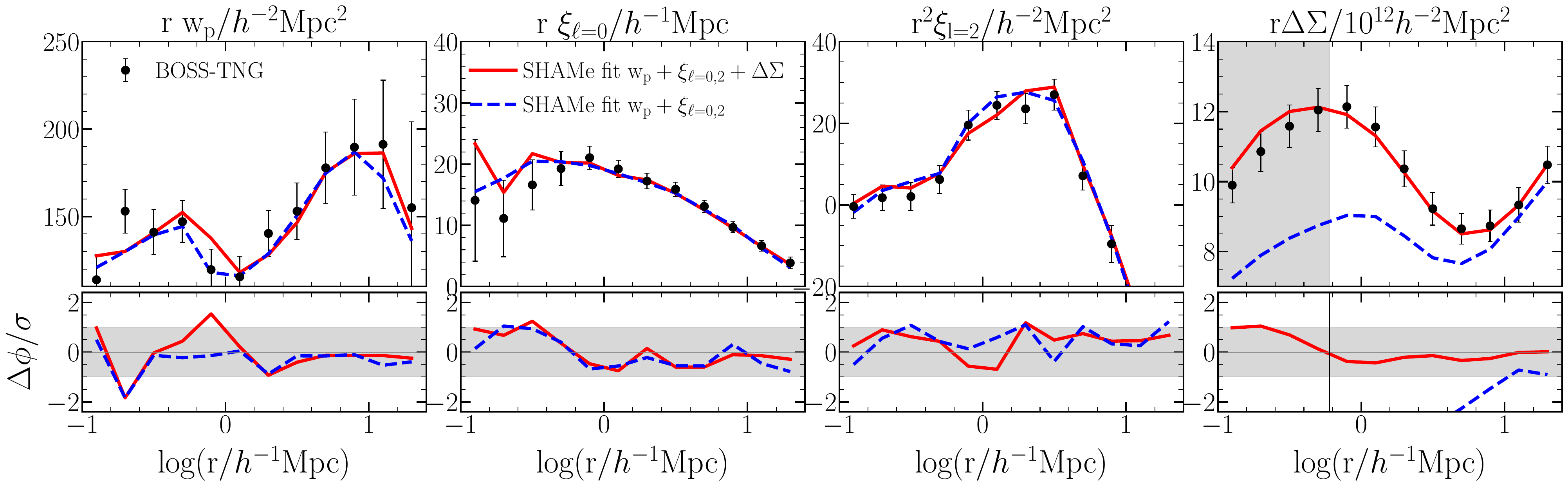}
\caption{\textbf{Top panel}. Projected correlation function (\proj), monopole (\mono), quadrupole (\quadr), and excess surface density (\lensing) for BOSS-TNG galaxies, a galaxy sample designed to reproduce the properties of BOSS-LOWZ galaxies. Dots and error bars indicate measurements and $1\sigma$ uncertainties for BOSS-TNG galaxies, while red and blue lines show the best-fitting SHAMe model to GC and GGL statistics (i.e., \proj+\mono+\quadr+\lensing) and GC (i.e., \proj+\mono+\quadr) of these galaxies, respectively. The grey area denotes the scales that were not utilised for the fitting of GGL. \textbf{Bottom panel}. The ratio between BOSS-TNG measurements and SHAMe predictions, normalise by the error in the measurements. The shaded area indicates the $1\sigma$ region.}  
\label{Fig:TNG_cluster}
\end{figure*} 

\begin{figure}
\includegraphics[width=0.45\textwidth]{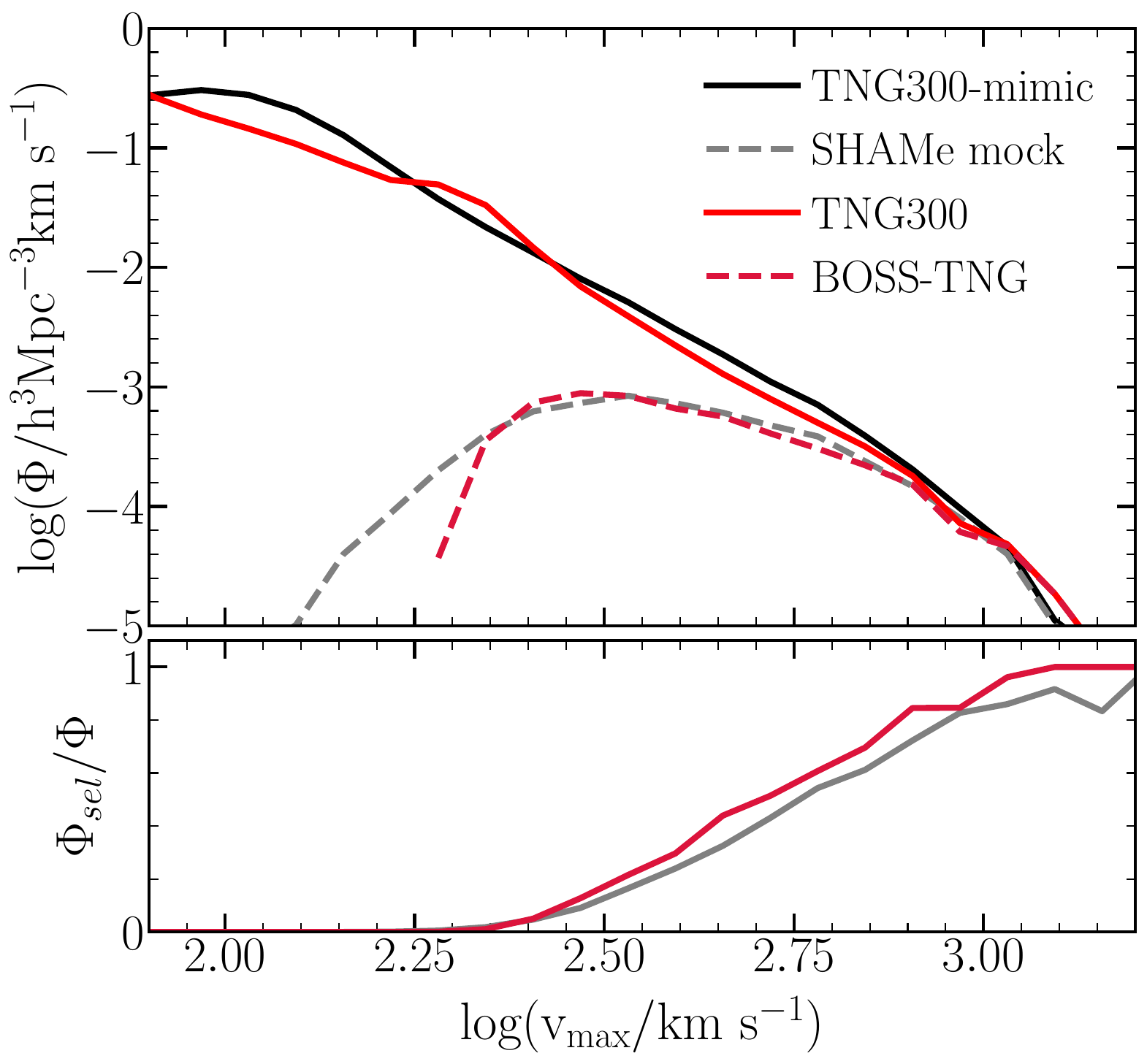}
\caption{ \textbf{Top panel}. Maximum circular velocity function for galaxies from the TNG300 simulation (solid red line), BOSS-TNG sample (dashed red line), TNG300-mimic subhalos (solid black line), and best-fitting SHAMe model to the GC and GGL of BOSS-TNG galaxies (dashed grey line). \textbf{Bottom panel}. The ratio between the completeness in $\vmax$ of the BOSS-TNG galaxy sample to the TNG300 galaxy catalogue and the SHAMe mock to the TNG300-mimic subhalo catalogue.}  
\label{Fig:TNG_coml}
\end{figure} 

While \cite{C23a} demonstrated that SHAMe can reproduce GC and GGL for $\rm M_r$-selected galaxies, \cite{Chaves:2023} found that the colour-based selection criteria of the BOSS survey are partially responsible for the lensing-is-low problem. Before applying SHAMe to observational data, we first test its ability to simultaneously reproduce the GC and GGL of a BOSS-like sample extracted from the TNG300 simulation \citep[see][for a detailed description]{Chaves:2023}.

To analyse this sample, SHAMe is evaluated on the TNG300-mimic simulation, which is a dark matter-only version of the TNG300 simulation with the same volume and initial conditions but a lower resolution ($625^3$ particles). Using the PSO algorithm described in \S~\ref{sec:PSO_MCMC}, we determine the SHAMe parameters that better described the GC and GGL of the BOSS-TNG galaxies. To achieve this, we use a covariance matrix computed from 64 distinct volumes of $\sim(350 \hMpc)^3$ taken from one of the ``Bacco Simulations'' \citep{Angulo:2021} with a volume of $1440 \hMpc$. Using the ``scaling technique''  \citep{Angulo:2021}, we scale this simulation to the \Planck~and \LowS~cosmology. Following \cite{C23a}, we performed a preliminary fit to the GC and GGL by assuming covariance matrices computed with jackknife errors from the hydrodynamic simulation \citep{Zehavi:2002, Norberg:2008}. Then, using the resulting best-fitting model, we populate the scaled ``Bacco Simulation'' to compute the final covariance matrix by using GC and GGL measurements from the 64 sub-volumes. Given that we do not anticipate any contribution from cosmic variances between the BOSS-TNG galaxies and the SHAMe mock due to the TNG300 and TNG300-mimic having the same initial conditions, this error is merely an estimate of the types of errors we would encounter in observations. In \cite{C23a}, we examine alternative methods for assigning errors to the TNG300 GC and GGL signals and find that they have a minimal effect on the SHAMe best-fitting parameter values.

In Fig.~\ref{Fig:TNG_cluster}, we show the GC (projected correlation function, and monopole and quadrupole of the correlation function) and GGL (excess surface density) of BOSS-TNG galaxies and the best-fitting SHAMe solutions to GC (blue lines) and both GC and GGL (red lines). We only fit the GGL signal down to 0.6 $\hMpc$ because baryonic effects become increasingly important for smaller scales. For scales above this threshold, the impact of baryons in the GGL is expected to be very limited (e.g., \citealt{Arico:2021b, Chaves:2023, C23a}). The first major finding of this study is that SHAMe simultaneously reproduces the GC and GGL of the BOSS-TNG sample, which was selected according to complex colour-based criteria. Therefore, we do not find any systematic difference between TNG300 measurements and SHAMe predictions for GGL, i.e., SHAMe does not suffer from a lensing-is-low problem for this sample. This result is quite encouraging as standard HOD models overpredict GGL for this sample \citep{Chaves:2023}, which motivates using SHAMe for interpreting observations.

We also find that the fit to GC is good regardless of whether GGL is included as constraining data, while GGL predictions vary significantly depending on whether this statistic is fitted or not. This result indicates that GC (GGL) imposes weak constraints on GGL (GC) for SHAMe, contrary to what is usually assumed in HOD studies when only fitting GC to predict GGL \citep{Leauthaud:2017, Lange:2019, Wibking:2020, Yuan:2022}.

BOSS galaxies have a completeness of only $\sim 80\%$ due to the complexity of the target selection \citep{Leauthaud:2016, Saito:2016}, which makes it challenging to model the GC and GGL of galaxies. To determine how well our model can mimic the completeness of a BOSS-like sample, we compare the completeness of the BOSS-TNG galaxy sample with the best-fitting SHAMe in Fig.~\ref{Fig:TNG_coml}. Specifically, we compare the $\vmax$ distribution of the first and second samples with that of the full galaxy and subhalo samples of the TNG300 and TNG300-mimic simulations, respectively. We used $\vmax$ since this property is a good proxy for the stellar mass of a galaxy \citep[e.g.,][]{ChavesMontero:2016} and is available in both simulations. 

We can readily see that both samples exhibit a similar $\vmax$ distribution, confirming the accuracy of the SHAMe model in recreating complex galaxy selections. While SHAMe does not include a free parameter dedicated to dealing with sample incompleteness, the combination of the parameters controlling the scatter in the SHAMe between the luminosity of the galaxies and $\vpeak$, assembly bias, and the abundance of satellites do so. It is important to note that we do not attempt to reproduce the completeness of the sample, this is just a prediction of the best-fitting model to GC and GGL. We also checked if adding a parameter for incompleteness improved the results, but we found that the performance of the model was similar.

%We repeated all these tests using a colour-selected sample from the MTNG simulation \citep{HernandezAguayo2022, Pakmor2022, Barrera2022, Kannan2022, Bose2022, Hadzhiyska2022a, Ferlito2022, Delgado2022, C23b}, finding similar results (not shown here).

\section{Predicting galaxy clustering and galaxy-galaxy lensing for BOSS galaxies}
\label{sec:cluster_BOSS}

\begin{figure*} 
\includegraphics[width=0.85\textwidth]{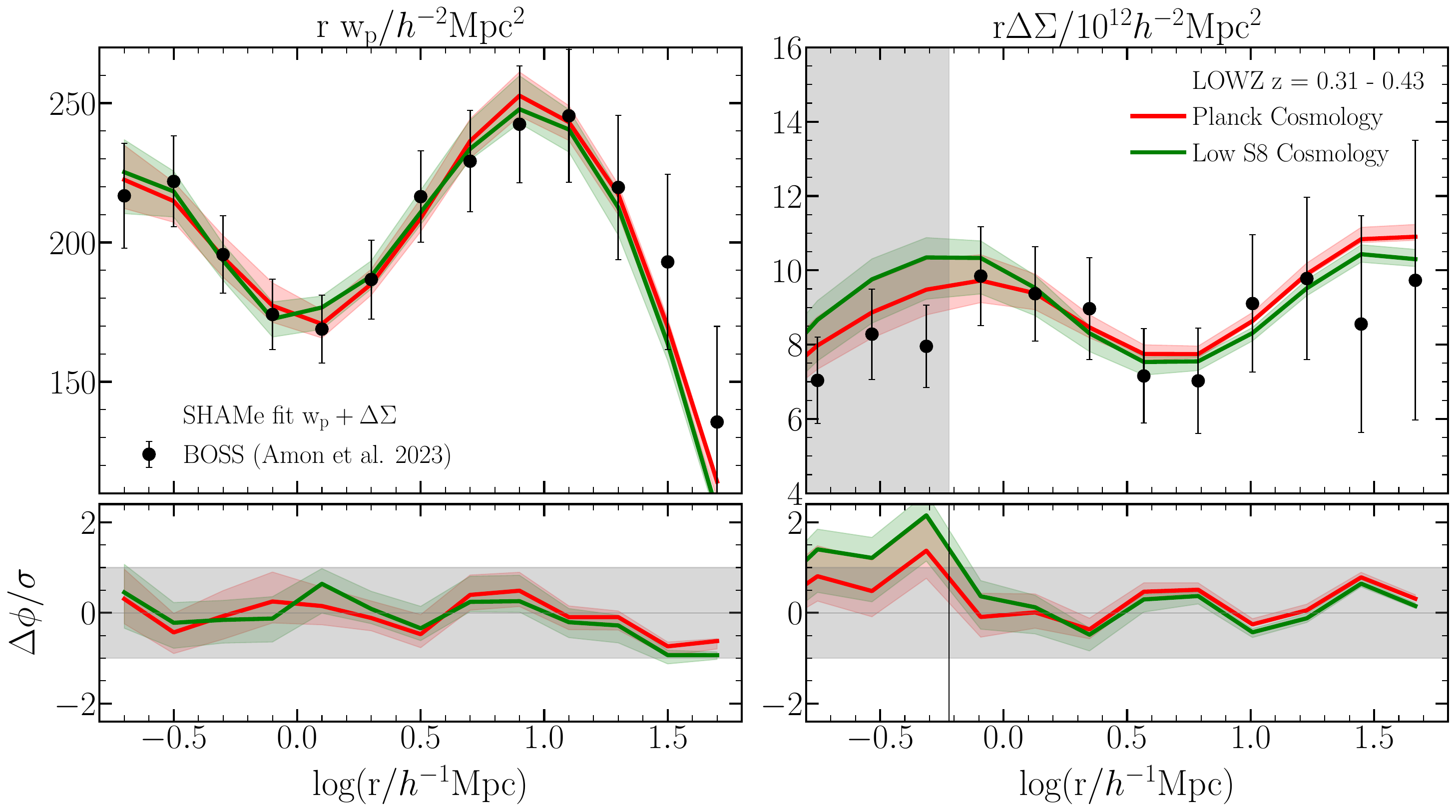}
\caption{Projected correlation function (left panel) and excess surface density (right panel) for LOWZ galaxies. Black dots indicate measurements from observations, while red and green lines denote best-fitting SHAMe models to GC and GGL adopting a \Planck~and \LowS~cosmology, respectively. The error bars and shaded areas denote 1-$\sigma$ uncertainties in the measurements and the predictions of the model. We do not use GGL measurements from scales smaller than 0.6 $\hMpc$ because these are affected by baryonic effects, which are not included in our model. The bottom panels display the ratio between measurements and model predictions normalised by observational uncertainties, with 1-$\sigma$ regions highlighted by gray-shaded regions.}
\label{Fig:BOSS_cluster}
\end{figure*}

\begin{figure}
\includegraphics[width=0.3\textwidth]{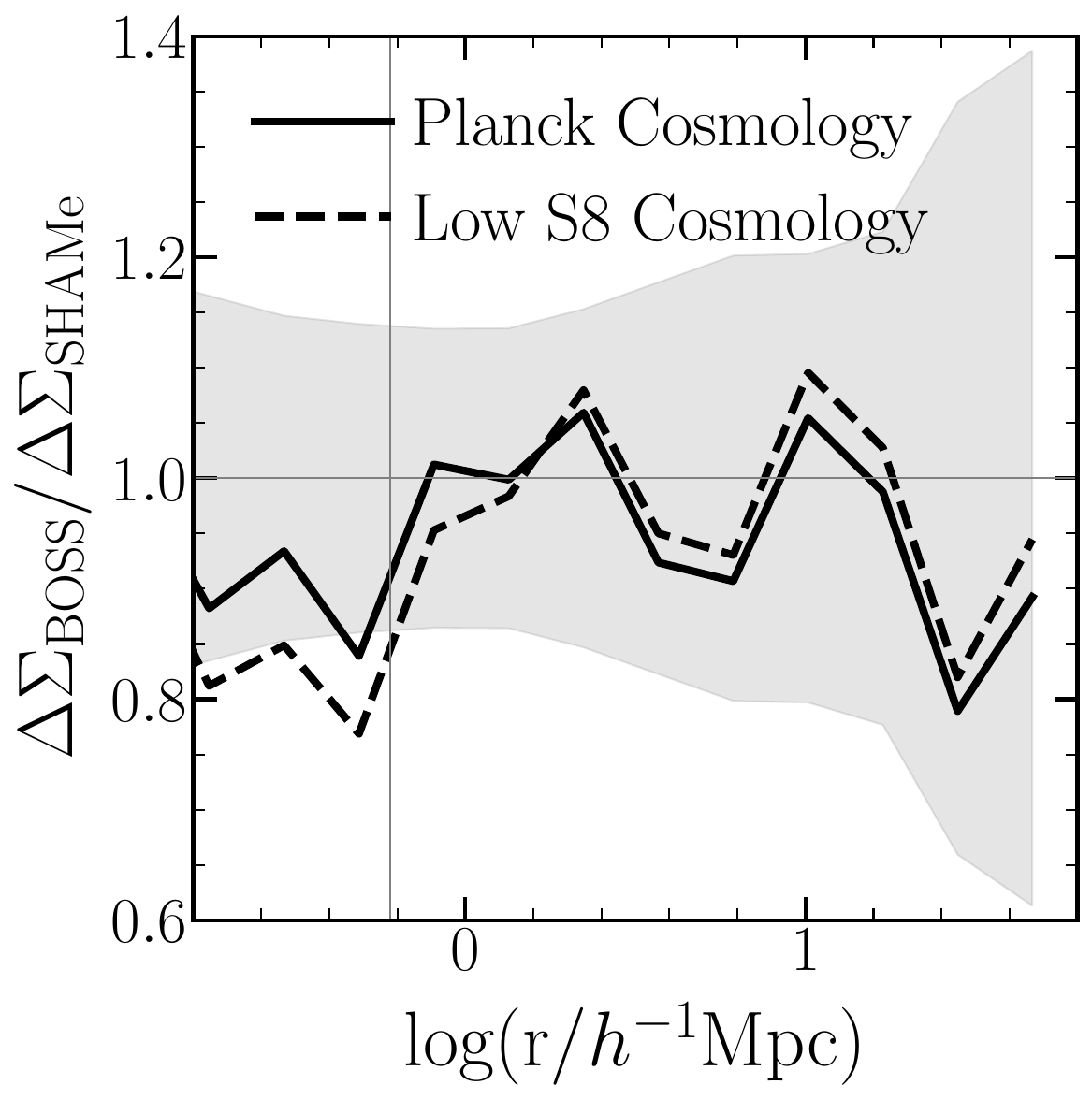} %0.3
\includegraphics[width=0.16\textwidth]{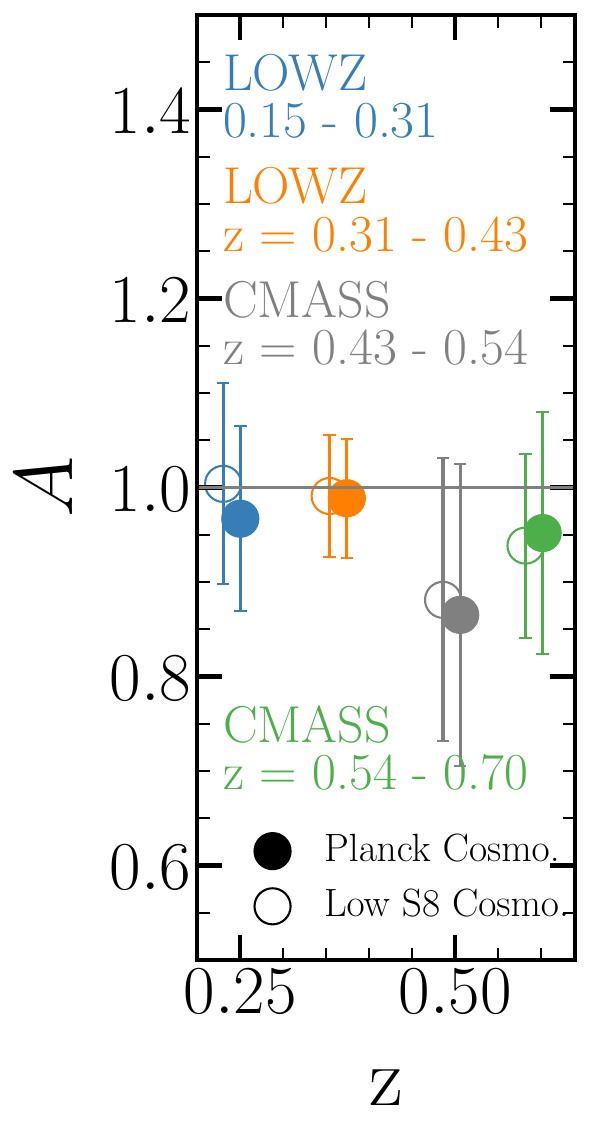} %0.15
\caption{\textbf{Left panel.} Ratio between excess surface density measurements and best-fitting model predictions. The solid and dashed line indicates the results when adopting the \Planck~and \LowS~cosmology, respectively. The grey-shaded region indicates 1-$\sigma$ uncertainties in observational measurements. SHAMe results in fits with no lensing-is-low for either cosmology, which would manifest as a strong systematic deviation of the lines from unity reaching $\sim 0.7-0.8$ on small scales. \textbf{Right panel.} Average departure of the ratio between GGL measurements and best-fitting models from unity for the four BOSS galaxy samples. As we can see, measurements and SHAMe predictions are consistent within 1-$\sigma$ uncertainties for all samples and cosmologies.}  
\label{Fig:Lensing_ratio}
\end{figure} 

\begin{figure}
\includegraphics[width=0.45\textwidth]{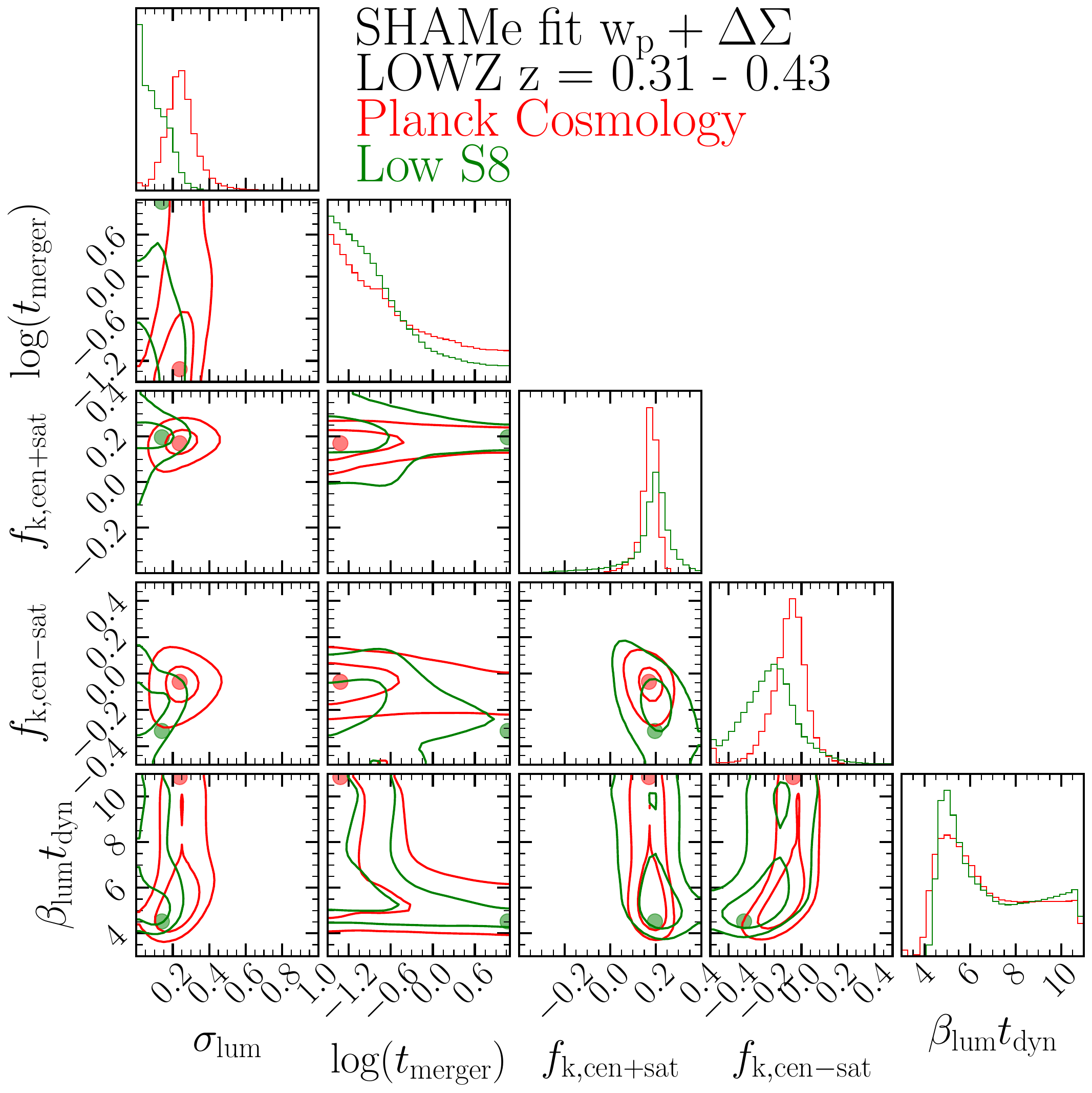}
\caption{Marginalised 1- and 2-$\sigma$ confidence levels for SHAMe parameters derived from fitting the GC and GGL of LOWZ galaxies. The red and green lines show the results for SHAMe when adopting the ``\Planck'' and `\LowS'' cosmology, respectively. Dots indicate best-fitting solutions.}
\label{Fig:Like_Planck}
\end{figure}

In the previous section, we show that SHAMe jointly reproduces GC and GGL for a simulated LOWZ-like galaxy sample. Motivated by this, in this section, we focus on reproducing these observables for galaxies from the LOWZ sample of the BOSS survey at $z=0.31-0.43$. We select this sample due to the excellent consistency between GGL measurements from different weak lensing surveys for these galaxies \citep{Leauthaud:2022}. In Appendix~\ref{sec:Ap1}, we show our findings for BOSS samples at higher and lower redshift.

In Fig.~\ref{Fig:BOSS_cluster}, we display the best-fitting solution of SHAMe to the GC and GGL of LOWZ galaxies, obtained using the emulator and MCMC framework described in \S~\ref{sec:shame} and \S~\ref{sec:PSO_MCMC}, respectively. The black symbols and error bars indicate observational measurements and uncertainties, while the red and green lines display best-fitting SHAMe predictions when we evaluate the model in a simulation adopting a ``\Planck'' and ``\LowS'' cosmology (see \S\ref{sec:simulations}), respectively. The shaded region represents the dispersion of the 3.7\% elements of the MCMC chains with the highest likelihood, equivalent to the area inside 1-sigma in a 5-dimension Gaussian (see \S~\ref{sec:PSO_MCMC} for more details). As for TNG300 data, we do not use GGL measurements on scales below 0.6 $\hMpc$ because our model does not incorporate baryonic effects. We can readily see that SHAMe accurately reproduces the GC and GGL of LOWZ galaxies for both cosmologies. We find a $\chi^2$ of 3.35 and 4.06 when adopting a ``\Planck'' and a ``\LowS'' cosmology, respectively, indicating that the precision of the model describing the data is very similar for both cosmologies. In addition, we can see that, even though we do not fit GGL on scales below 0.6 $\hMpc$, there is no significant difference between measurements and model predictions for these scales. Using the baryonification model of \cite{Arico:2021b}, we checked that fitting all scales shown while incorporating the baryonic effects predicted by the BAHAMAS simulation \citep{McCarthy:2017}, which are consistent with those measured in lensing data \citep{Arico:2023}, does not significantly change the results.

In the left panel of Fig.~\ref{Fig:Lensing_ratio}, we focus on the ratio between GGL measurements from BOSS and predictions from SHAMe for the two cosmologies, showing again that it is compatible with unity for all scales used in the fit within observational uncertainties. In the right panel of Fig.~\ref{Fig:Lensing_ratio}, we show the weighted average of the previous ratio for four different samples of the BOSS survey (two LOWZ and two CMASS samples), which we compute using \citep{Chaves:2023},
\begin{equation}
    \label{eq:lil_a}
    A \equiv 
    \dfrac{
    \sum_i \sigma^{-2}_i \Delta\Sigma_{\rm BOSS}(r_{i})/\Delta\Sigma_{\rm SHAMe}(r_{i})}{\sum_i \sigma^{-2}_i},
\end{equation}
where $i$ goes through all radial bins used in the analysis (> 0.6 $\hMpc$) and $\sigma_i$ is the uncertainty in the ratio $\Delta\Sigma_{\rm BOSS}/\Delta\Sigma_{\rm SHAMe}$ at a scale $r_i$. We estimate the error on $A$ by computing the weighted standard deviation of the lensing ratio. As we can see, all BOSS samples have a value of $A$ consistent with unity independently of the cosmology used to generate SHAMe predictions. We also find consistent results when computing $A$ using an unweighted mean instead of Eq.~\ref{eq:lil_a}. We can thus conclude that SHAMe does not encounter a lensing-is-low problem for LOWZ galaxies, confirming our expectations from the analysis of simulated LOWZ-like galaxies from the TNG300 simulation (see \S\ref{sec:cluster_TNG}). In addition, these results confirm that the lensing-is-low problem is primarily, if not entirely, a limitation of the way empirical galaxy-halo models model GC and GGL \citep{Chaves:2023}.

In Fig.~\ref{Fig:Like_Planck}, we show the constraints on SHAMe parameters when evaluating the model in simulations assuming a ``\Planck'' (red) and a ``\LowS'' (green) cosmology. The inner and outer contour lines indicate the 1- and 2-$\sigma$ confidence regions of the distribution, while the dots denote the best-fitting value of the parameters. As we can see, the parameters controlling the properties of satellite galaxies ($\tmerger$ and $\betaL$) present similar values for both cosmologies, indicating that the effect of these parameters on GC plus GGL is not degenerate with cosmology. On the other hand, the parameters controlling the scatter between galaxy luminosity and the depth of halo potential wells ($\sigL$) and assembly bias ($\sigL$, $\FkP$, $\FkM$) show a significant dependence on cosmology, which is expected given the degeneracy between the dependence of the amplitude of GC and GGL on these parameters and the amplitude of matter fluctuations ($\sigma_8$). We can see that the scatter between galaxy luminosity and the depth of halo potential wells is smaller for the ``\LowS'' cosmology than for the ``\Planck'' cosmology, which at fixed cosmology would translate into a greater amplitude of GC and GGL for the former, but that results in a similar amplitude for these observables due to the lower value of $\sigma_8$ for the second cosmology. Given the excellent agreement between GC and GGL measurements and best-fitting predictions for both cosmologies, we can conclude that the precision of these observables for LOWZ galaxies is not enough to discriminate between the ``\Planck'' and ``\LowS'' cosmology using SHAMe.

\section{Other summary statistics}
\label{sec:HOD_GAB}

\begin{figure*}
\includegraphics[width=0.85\textwidth]{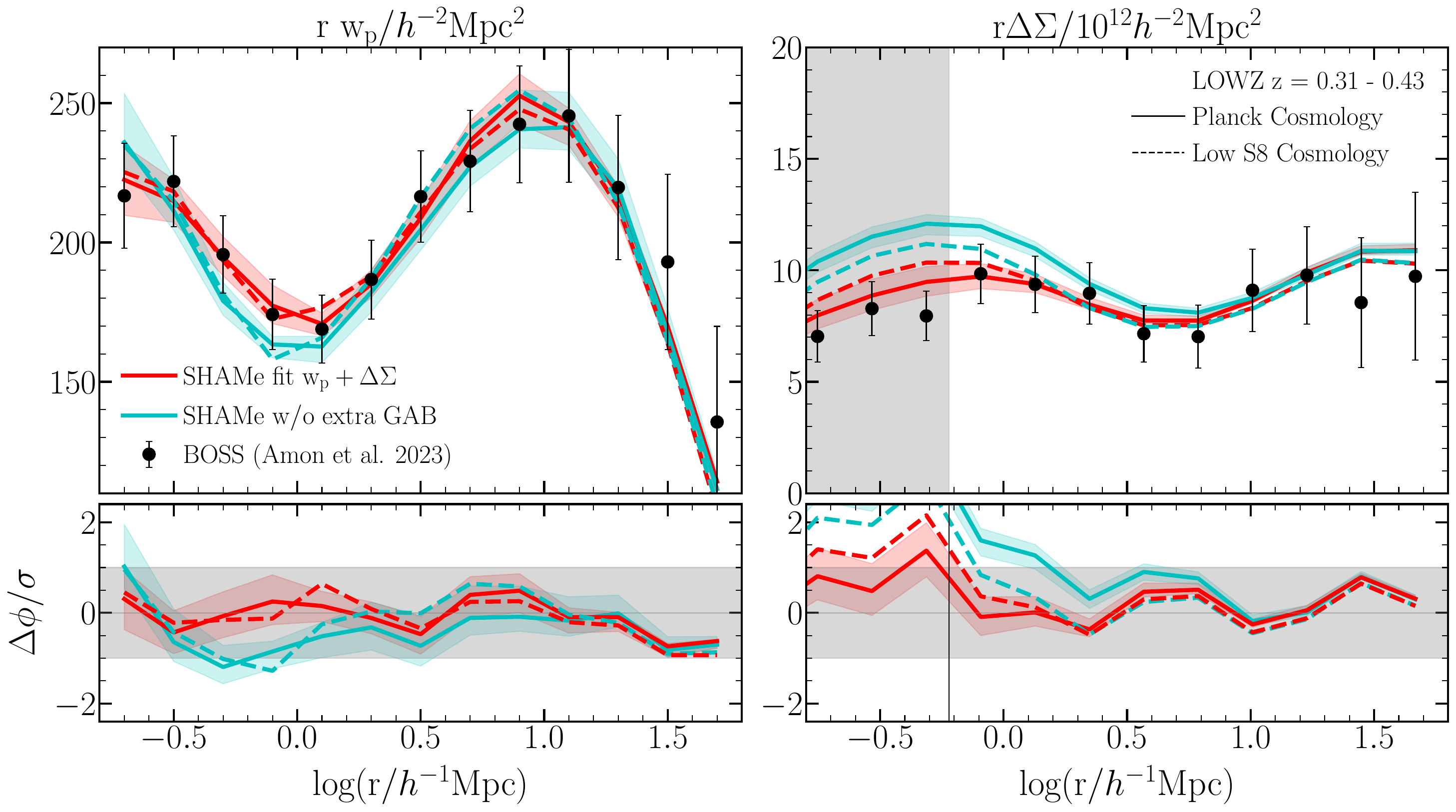}
\caption{{\bf Left panel.} Impact of galaxy assembly bias on SHAMe predictions for the LOWZ sample. Solid and dashed lines indicate the results for SHAMe when adopting the ``\Planck'' and ``\LowS'' cosmology, respectively, while red and cyan lines denote predictions with and without introducing additional galaxy assembly bias relative to a standard SHAM model. We use similar colour coding as in Fig.~\ref{Fig:BOSS_cluster}. The parameters of the model without assembly bias are obtained by re-fitting the GC and GGL. {\bf Right panel.} Similar to the top panel, but for SHAMe models that fit the GC only.}  
\label{Fig:GAB_clustering}
\end{figure*} 

\begin{figure}
\includegraphics[width=0.45\textwidth]{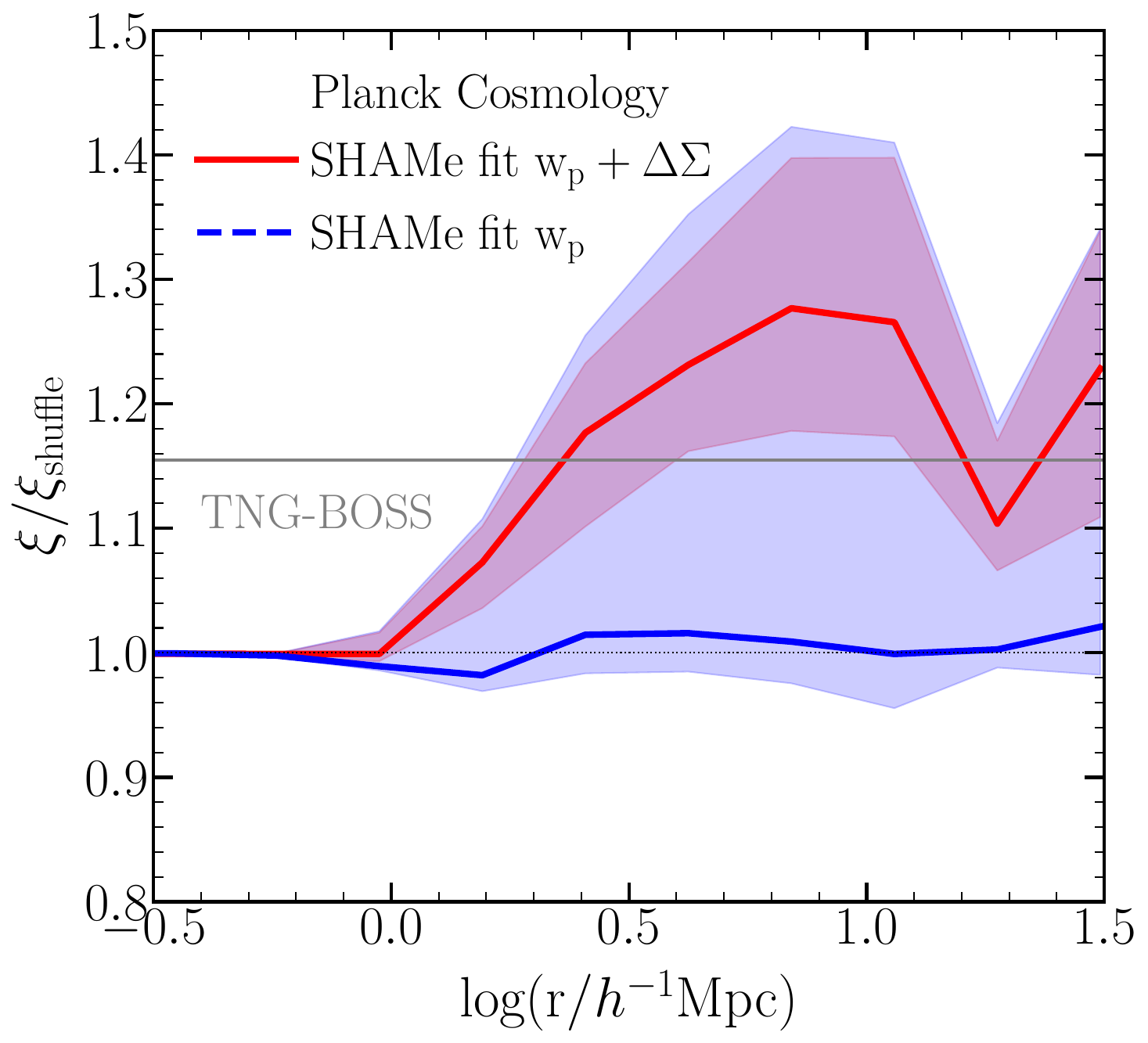}
\includegraphics[width=0.45\textwidth]{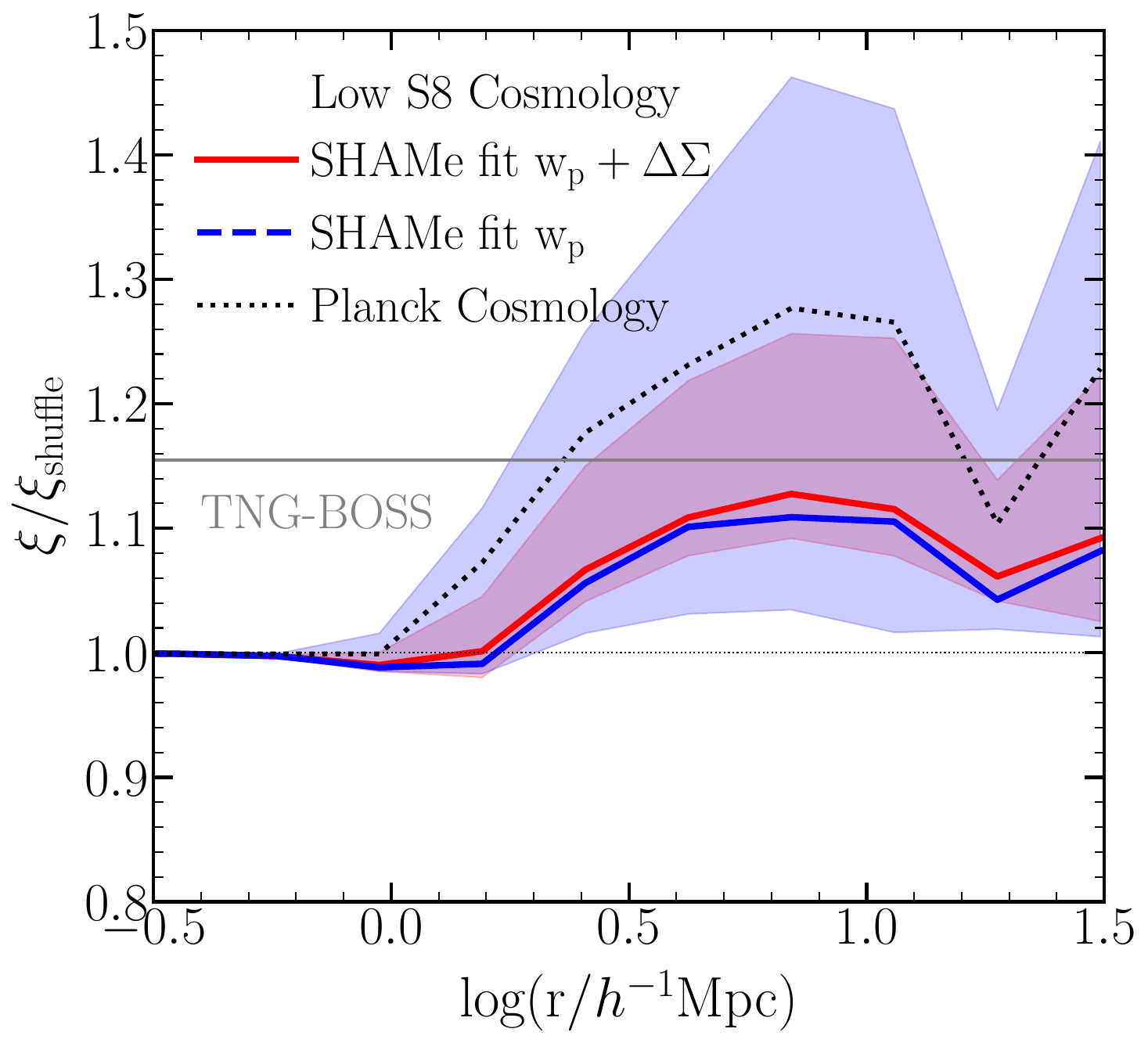}
 
\caption{
Galaxy assembly bias level for best-fitting SHAMe model to the LOWZ sample when adopting the ``\Planck'' cosmology {\bf (top panel)} and ``\LowS'' cosmology {\bf (bottom panel)}. Red lines show the results when using GC and GGL as constraining data, while blue lines do so when only using GC. Shaded areas indicate 1-$\sigma$ regions, while horizontal grey lines denote the level of galaxy assembly for the BOSS-TNG sample. In the {\bf bottom} panel, for a more straightforward comparison, the black dotted line displays the best-fitting SHAMe models to GC and GGL when assuming a ``\Planck'' cosmology.
}  
\label{Fig:gab} 
\end{figure} 

In this section, we extract constraints on the galaxy-halo connection from the best-fitting SHAMe models to LOWZ galaxies (see previous section), specifically on their level of galaxy assembly bias in \S\ref{sec:gab} and halo occupation distribution in \S\ref{sec:hod}. From the practical point of view, instead of using an emulator, we extract these constraints by first populating a gravity-only simulation using SHAMe, and then directly measuring the statistics of interest.

\subsection{Assembly bias}
\label{sec:gab}

Halo assembly bias is the dependence of the clustering of dark matter haloes on any halo property other than its mass \citep{Sheth:2004, Gao:2005, Gao:2007}. The number of galaxies per halo also depends upon some of the properties that result in halo assembly bias, such as concentration, the number of substructures, or halo age (among others), which is known as occupancy variation \citep{Zehavi:2018, Artale:2018}. The combination of both effects leads to a systematic bias in galaxy clustering, which is known as galaxy assembly bias, and it is typically measured by comparing the clustering of a galaxy sample with that of an identical galaxy population in which the positions of galaxies within haloes of similar mass are shuffled while maintaining the relative distance of the galaxies to the centre of their haloes \citep{Croton:2007}. This procedure removes any dependence of galaxy clustering on halo properties besides halo mass, and results in a 10-20\% systematic difference in the 2-point correlation function for stellar mass-selected galaxies on scales above $\sim 2 \hMpc$ ($\xi/\xi_{\rm shuffle}\sim 1.1-1.2$) for semi-analytical models \citep{Croton:2007}. This amount is comparable to that predicted by hydrodynamic simulations (e.g., \citealt{ChavesMontero:2016}) and slightly greater than that predicted by SHAMs (e.g., \citealt{C21a}), with a value that depends on the redshift, number density, and colour of the sample.

In the previous section, we computed the best-fitting SHAMe model to the GC and GGL of BOSS galaxies while leaving free the amount of galaxy assembly bias. To determine the impact of this effect on the fits, in Fig.~\ref{Fig:GAB_clustering} we compare these constraints with those extracted while holding fixed the level of assembly bias to that of a standard SHAM model (see \S\ref{sec:shame}). As we can see, when holding fixed the level of galaxy assembly bias, the quality of the GC and GGL fits degrades significantly for both cosmologies, and the level of GGL is overpredicted by $\sim 25$\% for the ``\Planck'' cosmology on small scales. Interestingly, this increase is consistent with the amplitude of the lensing-is-low problem found for galaxy-halo connection models not presenting flexible levels of galaxy assembly bias \citep[e.g.,][]{Leauthaud:2017}, and consistent with the impact of this effect on the lensing-is-low problem for BOSS-like galaxies in the TNG300 simulation \citep{Chaves:2023}. Nevertheless, as also shown in \cite{Chaves:2023}, there are a large variety of other galaxy formation effects that also need to be modelled to avoid the lensing-is-low problem, such as segregation of satellite galaxies relative to dark matter. This is likely why other attempts to solve the lensing-is-low problem by considering assembly bias have led to non-consistent results \citep{Yuan:2020, Yuan:2021, Yuan:2022, Yuan:2022b}.

Having established the importance of galaxy assembly bias to simultaneously fit GC and GGL, we display the level of this effect for the LOWZ sample in Fig.~\ref{Fig:gab}, which we extract using the aforementioned shuffling technique. The assembly bias constraints improve by a factor from 2 to 3 when adding GGL to GC as constraining data, highlighting the constraining power of this observable for assembly bias, in agreement with the increase in the constraints on galaxy assembly bias found in \citealt{C23a}. We also find that the best-fitting level of assembly bias is $\simeq10\%$ for the ``\LowS'' cosmology when using GC or both GC and GGL as constraining data, but it increases from practically zero to $\simeq20\%$ for the ``\Planck'' cosmology. Nevertheless, these two solutions are consistent within error bars.

For reference, black solid lines display the level of galaxy assembly bias for the BOSS-TNG sample ($\simeq 15\%$). Even though this sample was not designed to match the level of galaxy assembly bias for LOWZ galaxies, we can readily see that this level agrees with SHAMe predictions for the ``\Planck'' and ``\LowS'' cosmology when using both GC and GGL as constraining data. We can thus conclude that SHAMe predicts a level of galaxy assembly bias for both cosmologies compatible with predictions from state-of-the-art cosmological simulations\footnote{Note that the BOSS-TNG sample was extracted from the TNG300 simulation at $z=0$, which adopts the ``\Planck'' cosmology, but the dependence of assembly bias on either cosmology \citep{C21b} or redshift between $z=0$ to $\sim0.35$ \citep{C19, C21a, Montero-Dorta:2021} is negligible.}.

Although the predicted level of galaxy assembly bias could be degenerated not only with cosmology but also observational errors or model shortcomings, \cite{C23a} showed that SHAMe recovers the right level of galaxy assembly bias in the TNG300 simulation when using GC and GGL as constraining data. Taken together with the excellent fits to GC and GGL, make our constraint on galaxy assembly bias one of the most robust up to this day.

\subsection{Halo occupation distribution}
\label{sec:hod}

\begin{figure*}
\includegraphics[width=0.33\textwidth]{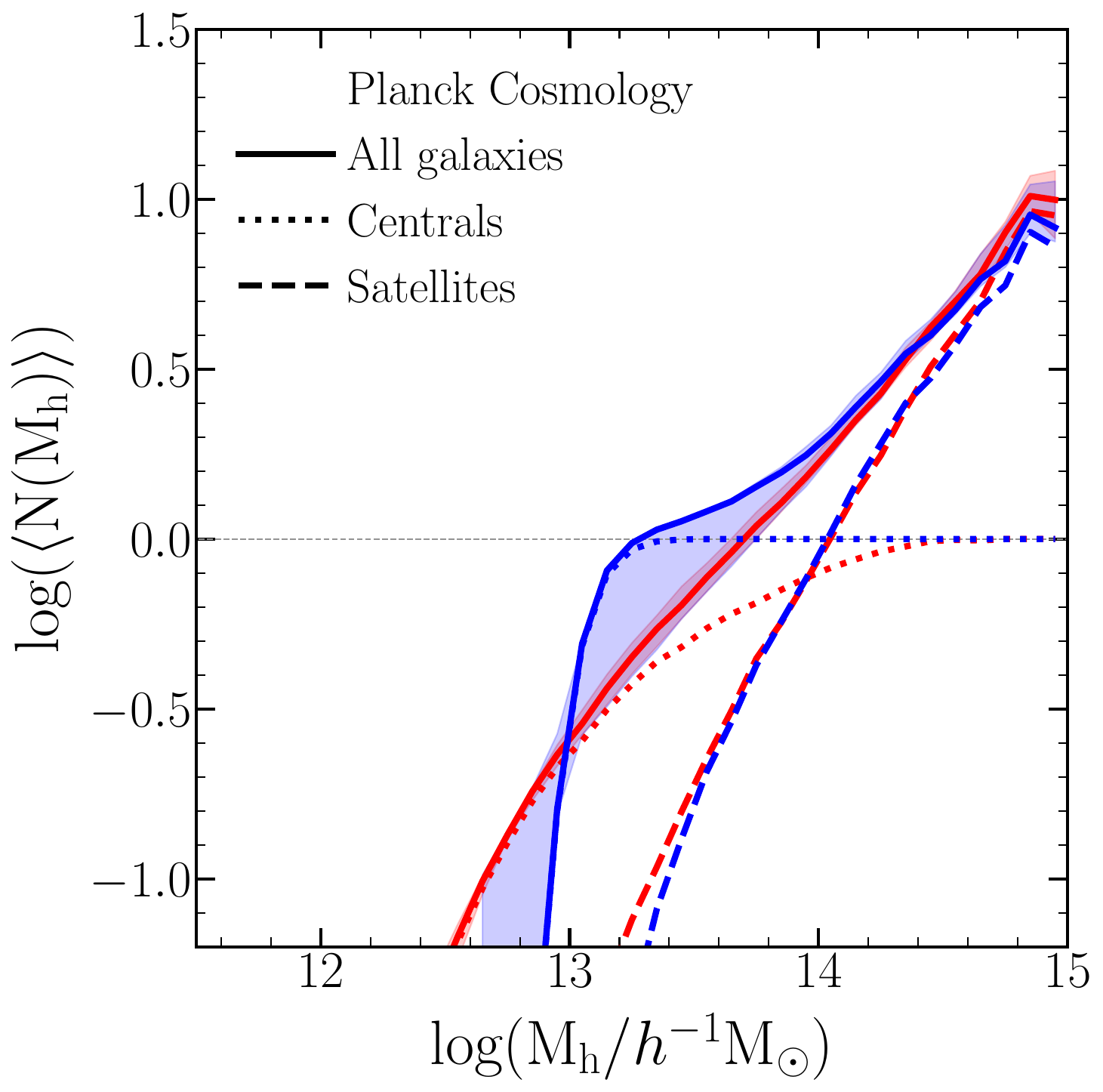}
\includegraphics[width=0.33\textwidth]{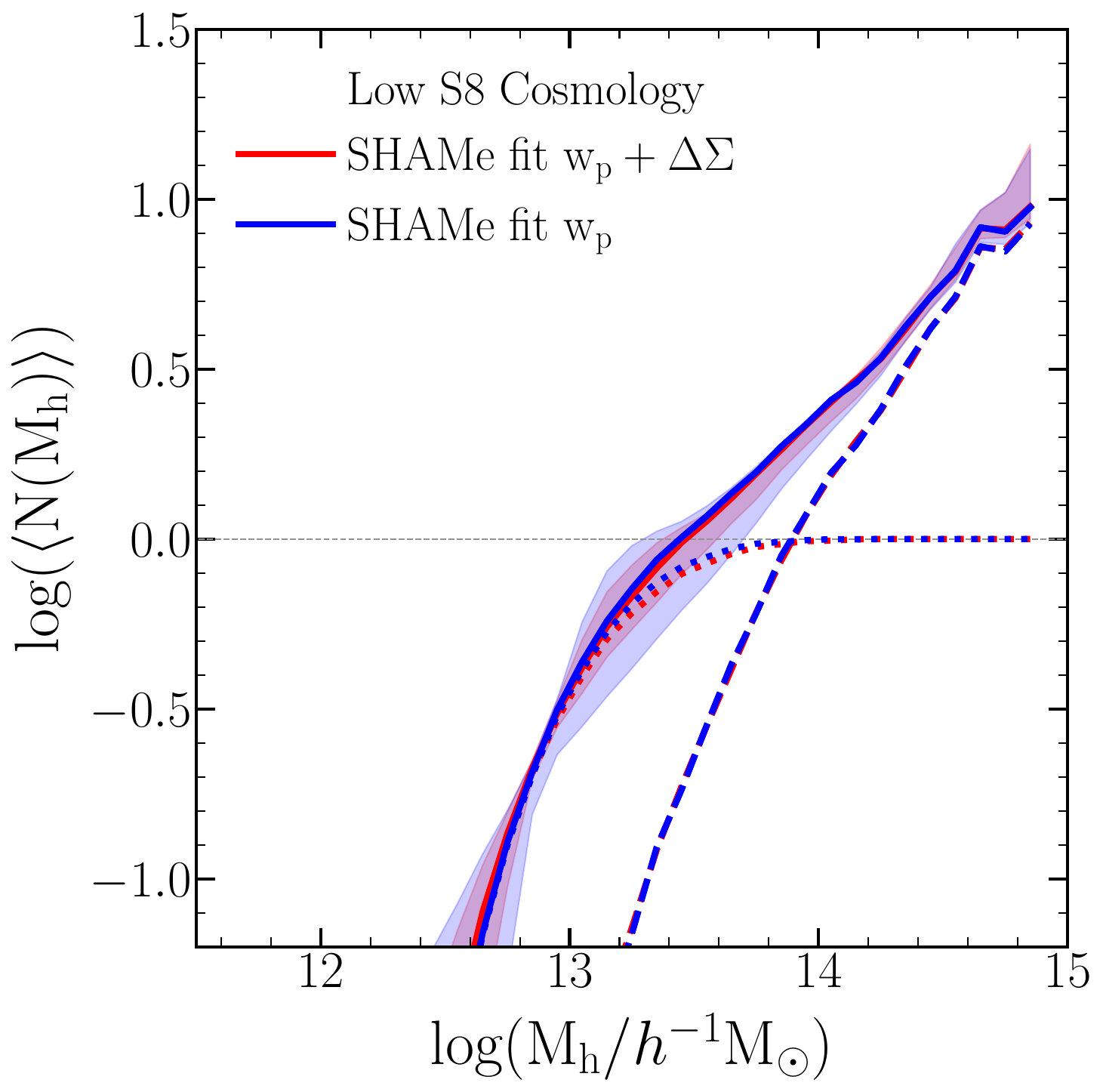}
\includegraphics[width=0.33\textwidth]{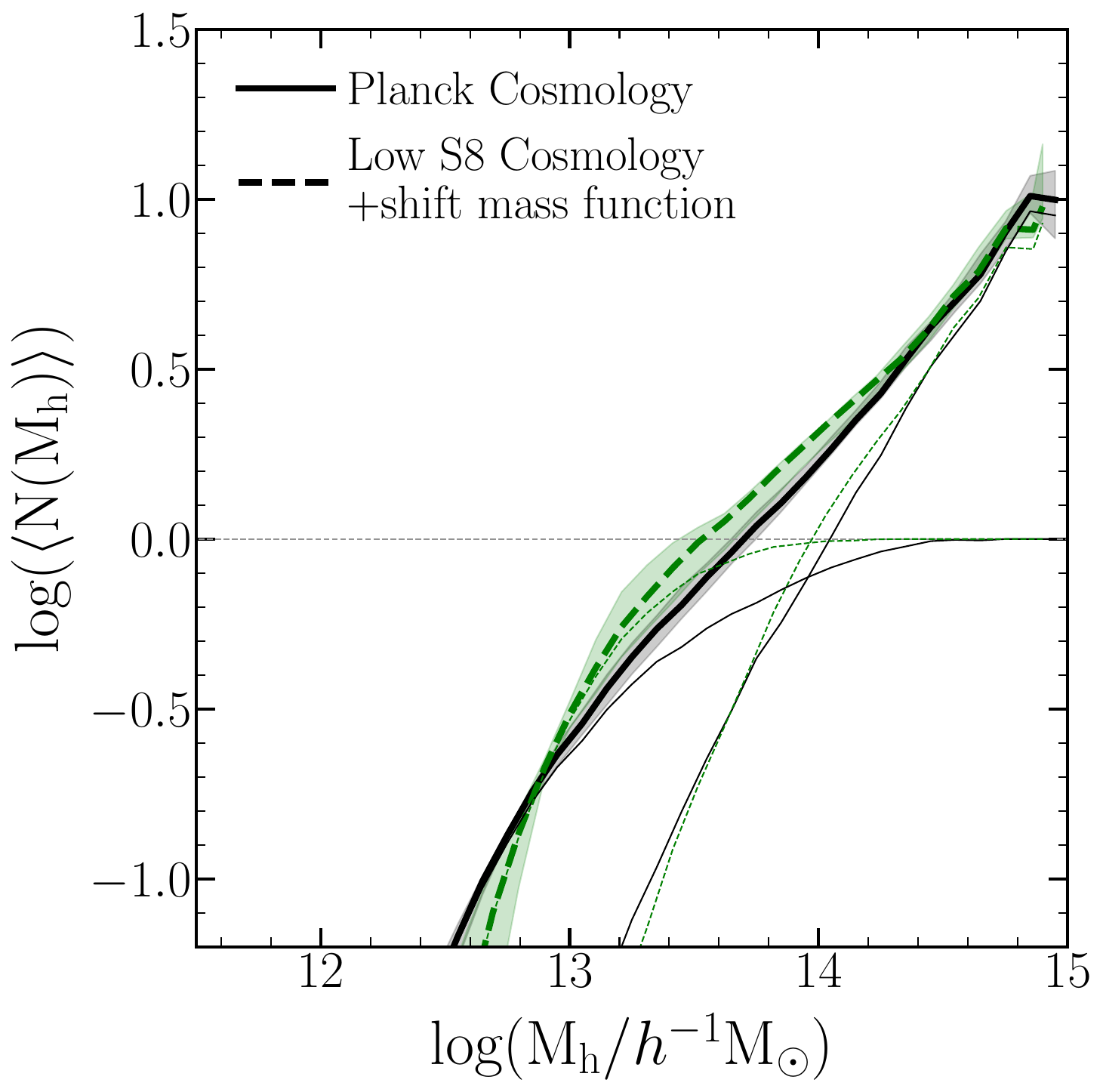}

\caption{Halo occupation distribution for best-fitting SHAMe model to the LOWZ sample when adopting a ``\Planck'' cosmology (left panel) and a ``\LowS'' cosmology (middle panel). Red lines show the results when using GC and GGL as constraining data, while blue lines do so when only using GC. Solid, dotted, and dashed lines display the contribution of all, central, and satellite galaxies, respectively, while shaded areas indicate 1-$\sigma$ regions. In the right panel, for a more straightforward comparison, the black solid and green dashed lines display the best-fitting SHAMe models to GC and GGL when assuming the ``\Planck'' and ``\LowS'' cosmology, respectively.}  
\label{Fig:hod}
\end{figure*}

In Fig.~\ref{Fig:hod}, we display the best-fitting halo occupation distributions for the LOWZ sample using a similar coding as in Fig.~\ref{Fig:BOSS_cluster}. Dotted, dashed, and solid lines indicate the results for central, satellites (including both galaxies in satellite subhalos and orphans), and all galaxies, respectively. We display the results for the ``\Planck'' and ``\LowS'' cosmology in the left and middle panels, respectively, and in the right panel, we show both for a more straightforward comparison. In this last panel, we modify halo masses so the halo mass function for the ``\LowS'' cosmology matches that for the ``\Planck'' cosmology. As we can see, including GGL improves the constraints on the width of the transition from zero to one central galaxy per halo (often represented by the $\sigmaLogM$ parameter in HOD models), which is typically poorly constrained when only using galaxy clustering as constraining data \citep[e.g.,][]{Zheng:2007, Zehavi:2011}. We can also see that the best fit for the ``Planck'' cosmology is at the edge of the 1-$\sigma$ region when only fitting GC. This is because the best-fitting value of $\sigL$ is very close to zero for this configuration, and this parameter approximately captures the aforementioned transition. 

We find that the best-fitting halo occupation distribution for the ``\LowS'' cosmology is practically the same when fitting just GC or both GC and GGL. However, we can readily see that the number of centrals in low-mass halos decreases for the \Planck~cosmology when including GGL in the fit. Interestingly, best-fitting SHAMe predictions for the \Planck~cosmology when using only GC as constraining present a lensing-is-low problem, and \citet{Chaves:2023} showed that one of the causes behind this problem is overpredicting the number of central galaxies in low-mass halos. In Appendix~\ref{sec:Ap2}, we fit a standard HOD parametric form to SHAMe predictions and provide the value of the resulting parameters.

\section{Conclusions}
\label{sec:fin}

In this paper we used the SHAMe model to simultaneously reproduce the galaxy clustering (GC) and galaxy-galaxy lensing (GGL) signal of galaxies from the BOSS survey. We proceed to summarize our most important findings:

\begin{itemize}

\item We test the performance of the SHAMe galaxy population model on the TNG300 hydrodynamic simulation, more specifically in a galaxy sample that resembles the LOWZ sample of the BOSS survey (BOSS-TNG, \citealt{Chaves:2023}). We find that our model successfully reproduces both the GC (\proj+\mono+\quadr) and GGL signal (\lensing) of these galaxies simultaneously (Fig.~\ref{Fig:TNG_cluster}). We also find that when fitting all these statistics, we can also reproduce the completeness in $\vmax$ of the BOSS-TNG galaxies (Fig.~\ref{Fig:TNG_coml}).

\item We find that SHAMe predictions for GGL present large uncertainties when only using GC as constraining data, letting us conclude that GC constraints on GGL are poor for our model. The precision of such predictions improves significantly when also considering GGL as constraining data, but it does not tighten GC predictions (see Figs.~\ref{Fig:TNG_cluster}~\&~\ref{Fig:BOSS_cluster_extra}).

\item Using the SHAMe model and two pairs of dark matter simulations, one with \Planck~cosmology and the other with a lower value of $\rm S8$ (``\LowS'' cosmology) we build several emulators capable of reproducing GC and GGL in a fraction of a second. Using these, we jointly reproduce the GC (\proj) and GGL (\lensing) of BOSS galaxies. Throughout the main body of this work, we focus on BOSS galaxies at $z=0.31-0.54$ (LOWZ sample) due to the great consistency among GGL measurements from different surveys for this sample (Fig.~\ref{Fig:BOSS_cluster}). We show the results for other BOSS samples in Appendix~\ref{sec:Ap1}.

\item We do not find systematic differences between GGL measurements and model predictions for either the ``\Planck'' or ``\LowS'' cosmology, i.e., SHAMe predictions are not affected by the lensing-is-low problem (Fig.~\ref{Fig:Lensing_ratio}).

\item We find that SHAMe predicts a non-negligible level of galaxy assembly bias for both the ``\Planck'' and ``\LowS'' cosmologies, $\simeq20$ and 10\%, respectively, which are both consistent with the level found for the BOSS-TNG sample ($\simeq 15\%$). We find that fitting both GC and GGL significantly tights galaxy assembly bias constraints, resulting in among the best constraints currently available (Fig.~\ref{Fig:gab}).

\item Joinly fitting GC and GGL also significantly improves halo occupation distribution constraints relative to only fitting GC, especially the width of the transition from zero to one central galaxy per halo, typically represented by the HOD parameter $\sigmaLogM$ (Fig.~\ref{Fig:hod}).

\end{itemize}

Throughout this work, we adopted two different cosmologies to generate SHAMe predictions: the first was compatible with early universe studies (``\Planck'' cosmology), while the second was in line with constraints from weak lensing surveys (``\LowS'' cosmology). We followed this approach because a popular interpretation of the lensing-is-low problem is that it is another face of the tension between the growth of structure measurements from the early and late universe \citep[e.g.,][]{Amon:2023}, and some works have found that this problem is alleviated when assuming a ``\LowS'' cosmology \citep{Leauthaud:2017, Lange:2019, Wibking:2020, Amon:2023, Yuan:2022}. However, we found no lensing-is-low problem for either the ``\Planck'' or ``\LowS'' cosmologies. In fact, the overall fit is better for the ``\Planck'' cosmology when considering only GC ($\chi^2_\mathrm{Planck} = 1.49$ vs $\chi^2_\mathrm{Low S8} = 2.77$) or both GC and GGL ($\chi^2_\mathrm{Planck} = 3.35$ vs $\chi^2_\mathrm{Low S8} = 4.06$) as constraining data. Note that these differences are not significant enough to conclude which cosmology is preferred by the data. The combination of the galaxy two-point correlation function ($\xi_{gg}$) and the cross-correlation between galaxies and dark matter ($\xi_{gm}$) has the potential to distinguish between two different cosmologies (i.e., $\xi_{gm}/\sqrt{\xi_{gg}} = \sqrt{\xi_{mm}}$). This implies that, in theory, the combined constraint of the projected correlation function (\proj) and galaxy-galaxy lensing (\lensing) should improve the constraints on $\sig$. Unfortunately, the observational errors, in this case, are too large to differentiate between these two cosmologies. Future GGL analysis on larger volumes will allow us to extract more information from the joint constraint using GC and GGL.

Independently of the performance of SHAMe for each cosmology, our main finding is that SHAMe provides accurate predictions for both GC and GGL, i.e., it is free from the lensing-is-low problem. \citet{Chaves:2023} suggested that this problem is caused by limitations of standard galaxy-halo connection models, which do not present enough flexibility to model the large variety of galaxy formation effects affecting GC and GGL. In \citet{C23a} and \S\ref{sec:cluster_TNG}, we showed that SHAMe in fact presents enough flexibility to account for most of these effects, which explains why we do not find a lensing-is-low problem for our model.

In a future project, we would like to test the performance of the model when for other clustering statistics such as the kNN-CDF \citep{Banerjee:2020}, bispectrum, and 3-point correlation function. We would also include the multipoles of the correlation function, which as shown in \cite{C23a}, can sustainably improve the constraints on SHAMe parameters over those already obtained using \proj\, and \lensing. Finally, we plan to combine the cosmological scaling of $N$-body simulations \citep{Angulo:2010, C20} with \shame to extract cosmological information from GC and GGL, following \cite{C23b}.

\section*{Data availability}

The IllustrisTNG simulations, including TNG300, are publicly available and accessible at \url{www.tng-project.org/data} \citep{Nelson:2019}. The data underlying this article will be shared on reasonable request to the corresponding author.

\section*{Acknowledgements}
We would like to thank Alexandra Amon and Naomi Robertson for kindly providing the data from their paper, \cite{Amon:2023} and Johannes Lange for useful comments.
SC acknowledges the support of the ``Juan de la Cierva Incorporac\'ion'' fellowship (IJC2020-045705-I).
REA and JCM acknowledge support from the ERC-StG number 716151 (BACCO). 
REA from the Project of excellence Prometeo/2020/085 from the Conselleria d'Innovaci\'o, Universitats, Ci\`encia i Societat Digital de la Generalitat Valenciana and JCM from the European Union's Horizon Europe research and innovation programme (COSMO-LYA, grant agreement 101044612). IFAE is partially funded by the CERCA program of the Generalitat de Catalunya.
The authors also acknowledge the computer resources at MareNostrum and the technical support provided by Barcelona Supercomputing Center (RES-AECT-2019-2-0012 \& RES-AECT-2020-3-0014)
Finally, we would like to thank our referee, David Weinberg, for his careful reading and helpful comments. % throughout the review process
\bibliographystyle{mnras}
\bibliography{Biblio} % if your bibtex file is called example.bib

\appendix

\section{The rest of the BOSS galaxy sample}
\label{sec:Ap1}

\begin{figure*}
\includegraphics[width=0.49\textwidth]{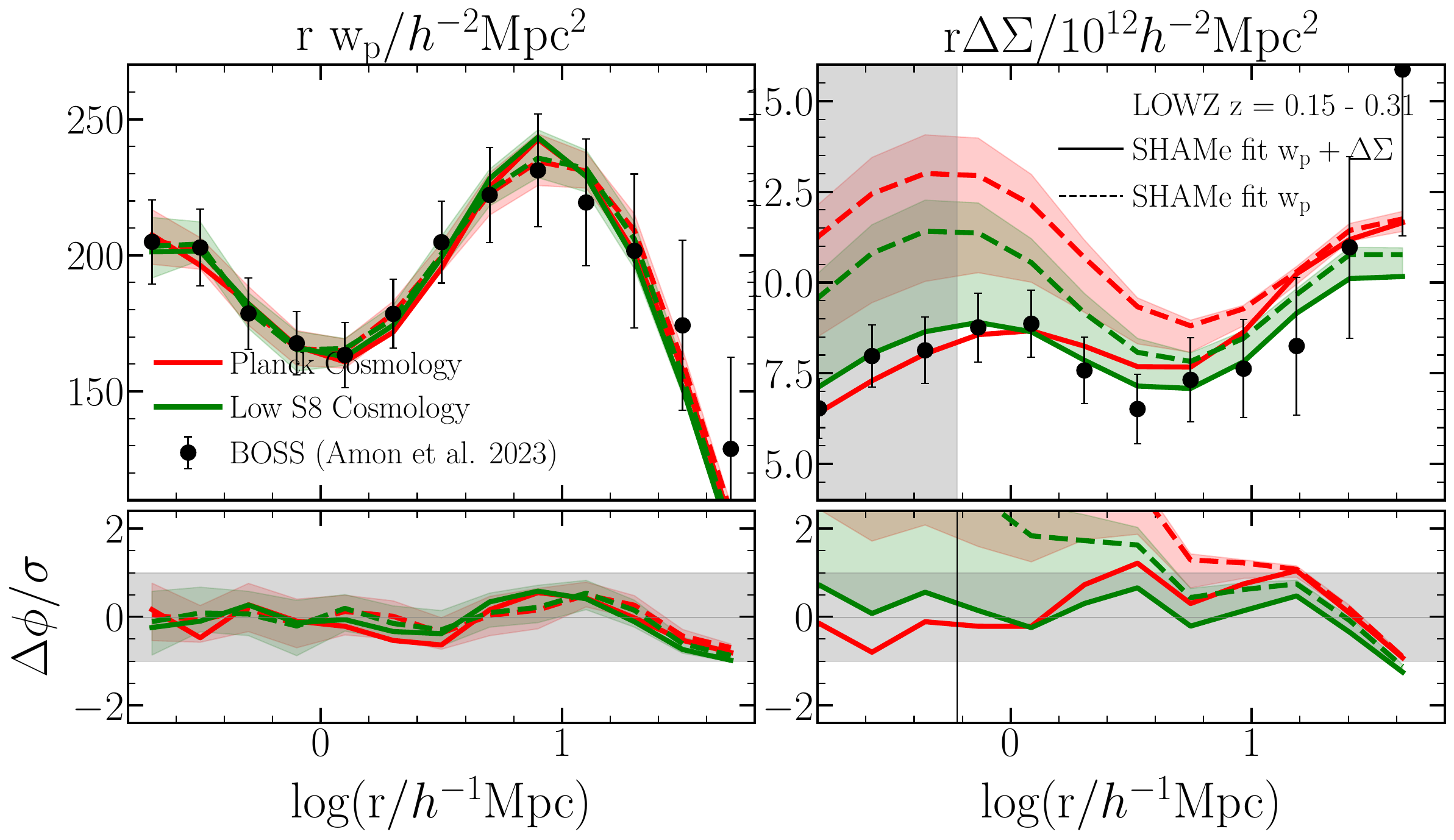}\includegraphics[width=0.49\textwidth]{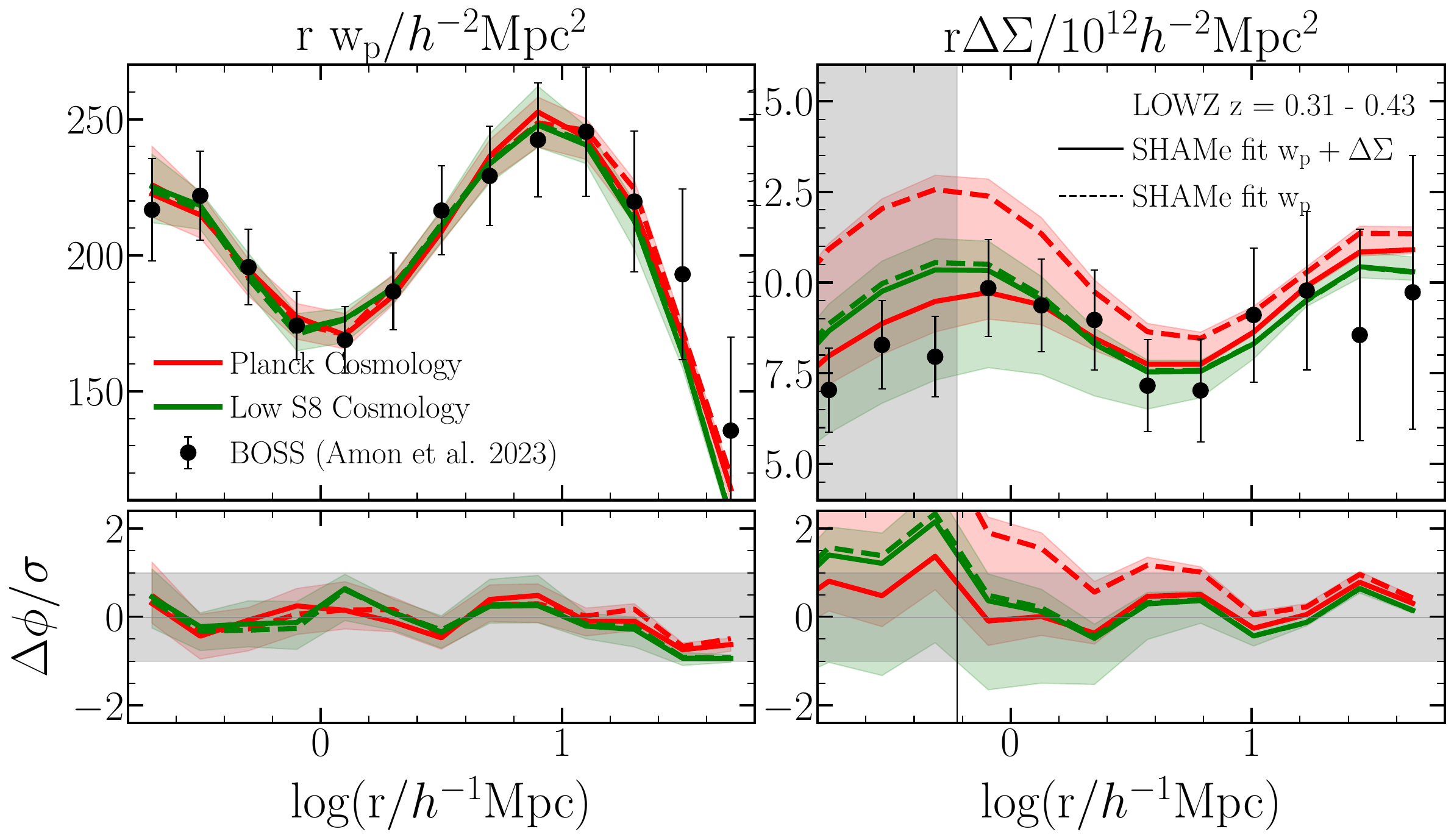}
\includegraphics[width=0.49\textwidth]{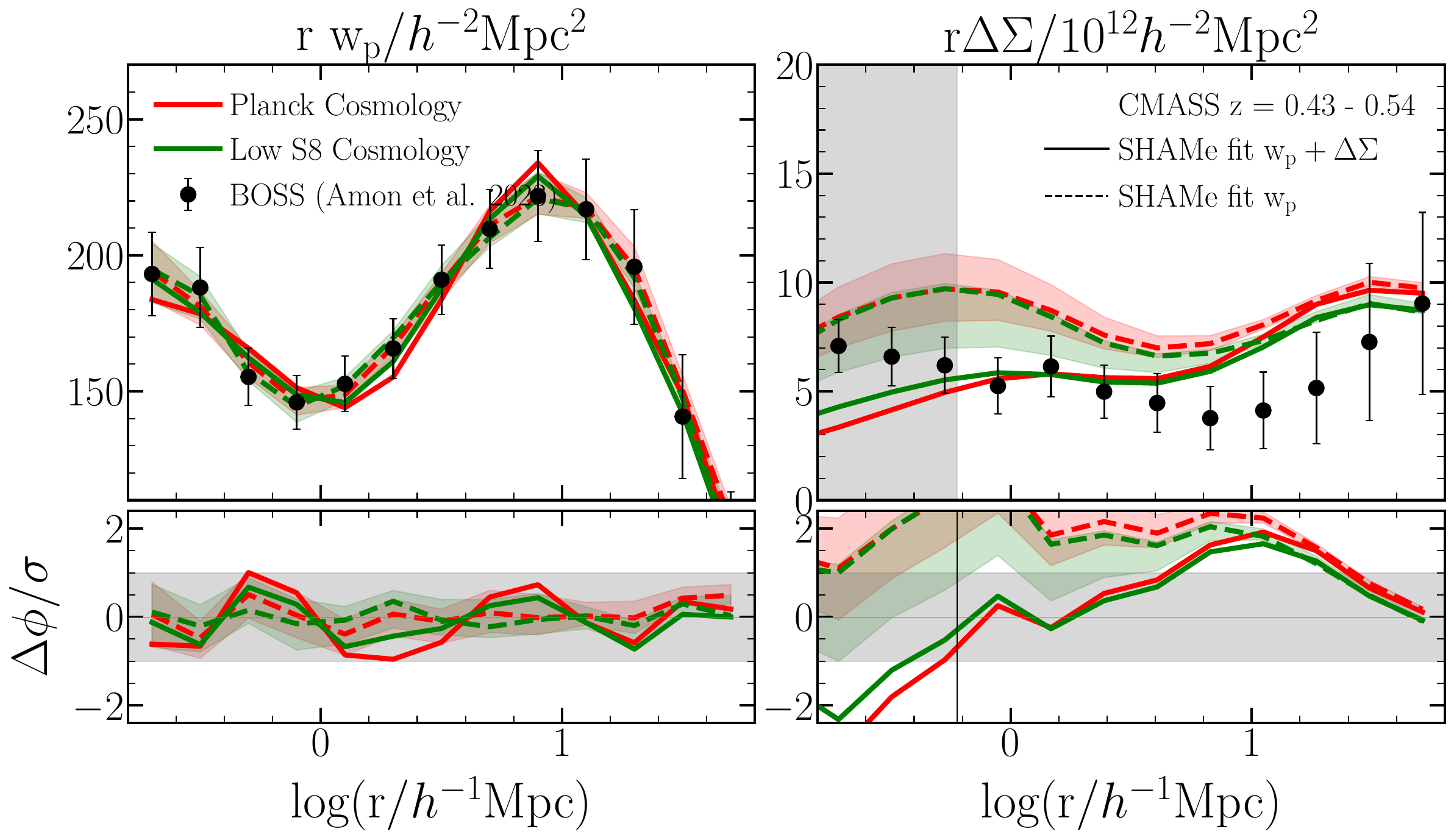}\includegraphics[width=0.49\textwidth]{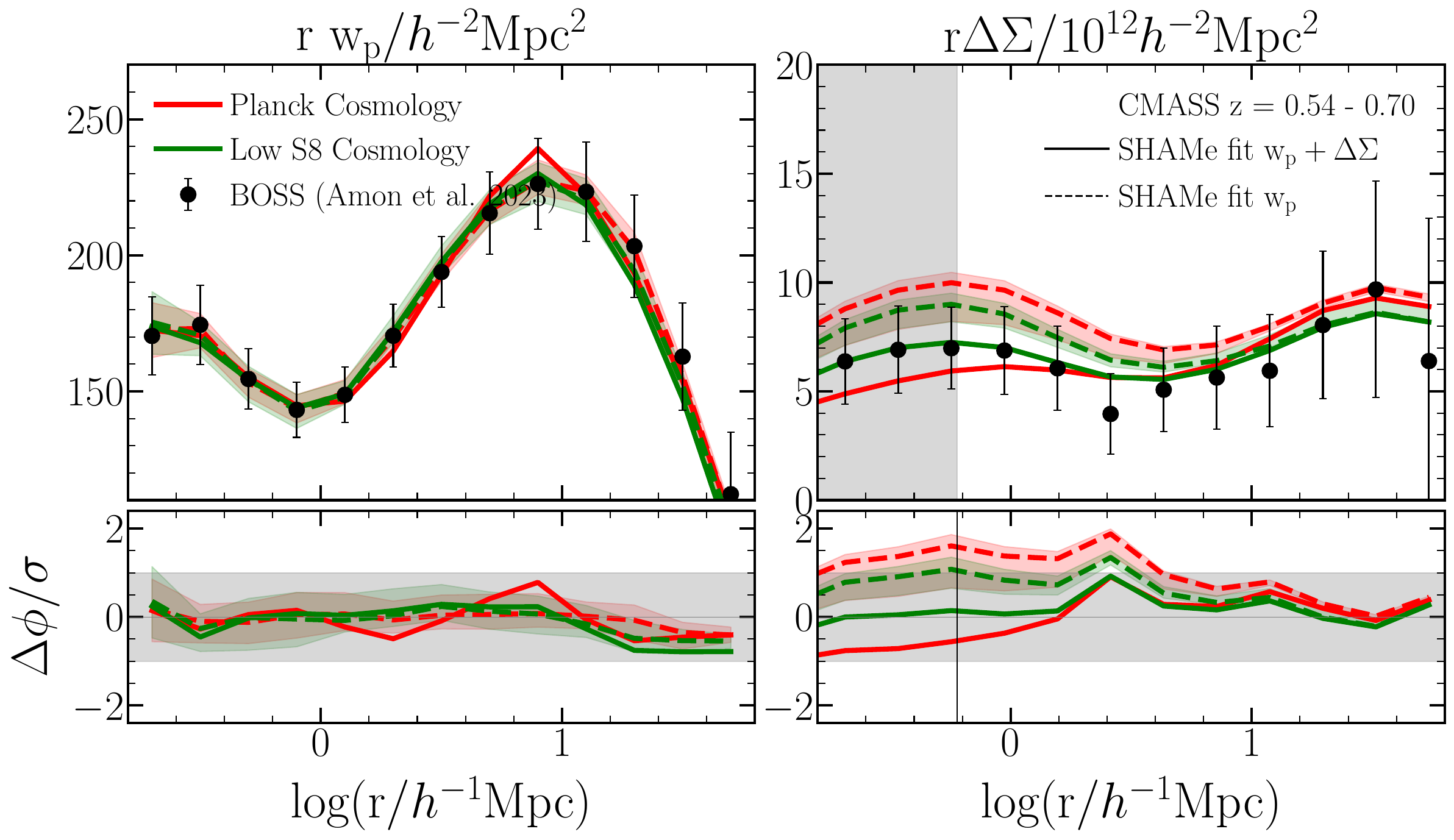}

\caption{Similar to Fig.~\ref{Fig:BOSS_cluster}, but for all BOSS samples from \citet{Amon:2023}: LOWZ at $z=0.15 - 0.31$ (top-left panel), LOWZ at $z=0.31 - 0.43$ (top-right panel), CMASS at $z=0.43 - 0.54$ (bottom-left panel), and CMASS at $z=0.54 - 0.7$ (bottom-right panel).}
\label{Fig:BOSS_cluster_extra}
\end{figure*} 

In this study, we focus on the LOWZ z = 0.31 - 0.43 galaxy sample from BOSS, primarily because it is the sample with the smallest discrepancy in the GGL signal between different observational studies \citep{Leauthaud:2022}. Despite this, we still calculate the GC and GGL from the remaining observational samples: LOWZ z = 0.15 - 0.31, CMASS z = 0.43 -0.54 and CMASS z = 0.54 - 0.70. Due to its performance, the CMASS z=0.43 - 0.54 was excluded from some analyses of \cite{Amon:2023}.

The GC and GGL measurements for these additional samples are shown in Fig.~\ref{Fig:BOSS_cluster_extra}. {Here we also include the fits when solely fitting GC (blue lines).} The shaded region corresponds to the 3.7\% of the elements of the MCMC chains with the highest likelihood of the fit with GC only. The dispersion when using GC and GGL is smaller, and similar between the two cosmologies, as shown in Fig.~\ref{Fig:BOSS_cluster}. We can reproduce the GC and GGL for all the samples except for CMASS z=0.43-0.54. While there is a systematic difference in the GGL signal for this last sample, this difference is not more pronounced at smaller scales; therefore, we would not classify it as a ``lensing-is-low'' signal, but rather as an inconsistency between the GC and GGL measurements. This extends the conclusion reached with the LOWZ z = 0.31 - 0.43 sample, namely that the SHAMe model can reproduce GC and GGL simultaneously and without the lensing-is-low problem.

\section{HOD fitting}
\label{sec:Ap2}
We fit the HODs computed in \S~\ref{sec:hod} using the 5-parameter formalism presented in \cite{Zheng:2005}, which characterizes the average halo occupation, $\langle N(M)\rangle$, by the halo mass at which half the haloes host a central galaxy above a given threshold ($\Mmin$), the width of the transition from zero to one galaxy per halo ($\sigmaLogM$), the halo mass where haloes start being occupied by satellite galaxies ($\Mcut$) the halo mass where there is on average one satellite galaxy per halo ($\Mone$) and the slope of the power law that characterise the increase of satellite galaxies as a function of halo mass ($\alpha$):
\begin{equation}
 \langle N_{\rm gal}(M_{\rm h})\rangle =  \langle N_{\rm cen}(M_{\rm h})\rangle +  \langle N_{\rm sat}(M_{\rm h})\rangle.
\end{equation}
with
\begin{equation}
 \langle N_{\rm sat}(M_{\rm h})\rangle = \left( \frac{M_{\rm h}-M_{\rm cut}}{M^*_1}\right)^\alpha,
\label{Eq:Sat_HOD}
\end{equation}
and 
\begin{equation}
 \langle N_{\rm cen}(M_{\rm h})\rangle = \frac{1}{2}\left[ 1 + {\rm erf} \left( \frac{\log M_{\rm h} - \log M_{\rm min}}{\sigma_{\log M}}  \right) \right],
\label{Eq:Cen_HOD}
\end{equation}
with $\rm erf(X) = \frac{2}{\sqrt{\pi}} \int_{0}^{x} e^{-t^2} {\rm d}t$ (the error function). We present the best-fitting parameters in Table~\ref{Table:HOD}. Please notice that a HOD mock build with these parameters should not reproduce the GC nor the GGL as we showed in this paper, since these are the HODs of the SHAMe mocks, and include all the additional complexity and performance of our empirical model, which are not normally present on a standard HOD. That being said, we believe the resulting HOD should represent better the target galaxy population since is not biased by the limitations of the HOD when fitting GC and GGL.

\begin{table}
    \centering
    \caption{The HOD parameters that best characterised the occupation number of the SHAMe models used to fit the LOWZ galaxy sample. The functional form assumed for the HOD is presented in equations 14 - 16 and corresponds to the Zheng et al. (2005)  HOD model. The errors correspond to the variation in HODs of the 3.7\% SHAMe mocks with the lowest $\chi^2$ from the MCMC chains when fitting the GC and/or GGL. }%\cite{Zheng:2005}
    \begin{tabular}{ccccc}
        \hline
        Params. & \Planck \proj & \proj+\lensing & \LowS \proj & \proj+\lensing\\
        \hline
        ${\rm log}(\Mmin)$ & $13.06^{+0.55}_{-0.00}$ & $13.46^{+0.14}_{-0.11}$ & $13.13^{+0.48}_{-0.10}$ & $13.15^{+0.14}_{-0.10}$ \\
        $\sigmaLogM$       &  $0.15^{+0.90}_{-0.00}$ &  $0.90^{+0.16}_{-0.14}$ &  $0.47^{+0.63}_{-0.23}$ &  $0.51^{+0.22}_{-0.18}$ \\
        ${\rm log}(\Mcut)$ &  $13.21^{+0.04}_{-0.42}$ &  $12.90^{+0.13}_{-0.19}$ &  $13.12^{+0.13}_{-0.23}$ &  $13.12^{+0.13}_{-0.20}$ \\
        ${\rm log}(\Mone)$ & $13.96^{+0.05}_{-0.03}$  &  $14.01^{+0.01}_{-0.04}$ &  $13.84^{+0.09}_{-0.02}$ &  $13.84^{+0.08}_{-0.01}$ \\
        $\alpha$           &  $1.02^{+0.17}_{-0.03}$&  $1.14^{+0.10}_{-0.07}$&  $1.06^{+0.15}_{-0.07}$&  $1.06^{+0.15}_{-0.05}$\\
        \hline
    \end{tabular}
\label{Table:HOD}
\end{table}

\bsp	% typesetting comment
\label{lastpage}
\end{document}